\documentclass[twocolumn]{jpsj2}

\title{%
Tricritical Behavior in Charge-Order System}

\author{%
Takahiro \textsc{Misawa}\thanks{E-mail:misawa@issp.u-tokyo.ac.jp},
Youhei \textsc{Yamaji} and Masatoshi \textsc{Imada}  
}

\inst{
 {\it Institute for Solid State Physics,
   University of Tokyo, 5-1-5, Kashiwanoha, Kashiwa, Chiba 277-8581}
}

\abst{Tricritical point in charge-order systems and its criticality 
 are studied for a microscopic model by using the mean-field approximation 
and exchange Monte Carlo method in the classical limit
as well as by using the Hartree-Fock approximation for the quantum model.
 We study the extended Hubbard model and show that the tricritical point 
emerges as an endpoint of the  first-order transition 
line between the disordered phase and the charge-ordered
phase at finite temperatures.
Strong divergences of several fluctuations at zero wavenumber
are found and analyzed around the tricritical point.
Especially, the charge susceptibility $\chi_{c}$ and 
the susceptibility of the next-nearest-neighbor correlation $\chi_{R}$
are shown to diverge and their critical
exponents are derived to be the same 
as the criticality of the susceptibility of the double 
occupancy $\chi_{D({\bf 0})}$. 
The singularity of conductivity at the tricritical point is clarified. 
We show that the singularity of the conductivity $\sigma$
is governed by that of the carrier density and is given as
$|\sigma-\sigma_{c}|\sim|g-g_{c}|^{p_{t}}(A\log{|g-g_{c}|}+B)$, where $g$ is 
the effective interaction of the Hubbard model, $\sigma_{c}$
($g_{c}$) represents the critical conductivity(interaction) 
and $A$ and $B$ are constants, respectively.
Here, in the canonical ensemble, we obtain $p_{t}=2\beta_{t}=1/2$  at the tricritical point.
We also show that $p_{t}$ changes into $p_{t}^{\prime}=2\beta=1$ at 
the tricritical point in the grand-canonical ensemble when the tricritical point
in the canonical ensemble is involved within the phase separation region.  
The results are compared with available experimental results of
organic conductor (DI-DCNQI)$_{2}$Ag. 
}

\kword{tricritical point, extended Hubbard model,
charge order, organic conductor,
charge susceptibility, doublon susceptibility, phase separation
}

\begin{document}
\maketitle

\section{Introduction}
Charge orderings and charge density waves are widely observed in various 
correlated electron systems such as transition metal compounds~\cite{RMP} 
and organic conductors~\cite{Seo}.
Such regular alignment of electrons with a periodicity longer than 
that of the unit cell at high temperatures is stably realized, 
particularly when the electron density is commensurate, where the number of electrons 
per unit cell is a simple fraction~\cite{Noda}. 
These ubiquitous phenomena have attracted recent intensive interest not 
only because of its own right but also because dramatic 
phenomena such as colossal magnetoresistance in perovskite-type manganese oxides
are observed immediately when the charge order melts. 
Charge orderings are in many cases 
competing or coexisting with magnetism, ferroelectricity and superconductivity, 
resulting in strong coupling to transport and magnetic properties.
Charge orderings are in fact frequently accompanied by metal-insulator transitions 
through the growth of charge-order parameter and the opening of gaps at the 
Fermi surface. 

Sensitive changes in transport, optical, dielectric and magnetic properties at or near
charge-order transitions have stimulated intensive research on their
control through the charge-order transitions.
To understand various possibilities, it is desired to clarify the basis of
charge-order transitions themselves, since the sensitive changes of physical properties 
and emergence of competing phases may be deeply  influenced by the nature of the transition.

\begin{figure}[htp]
	\begin{center}
		\includegraphics[width=8cm]{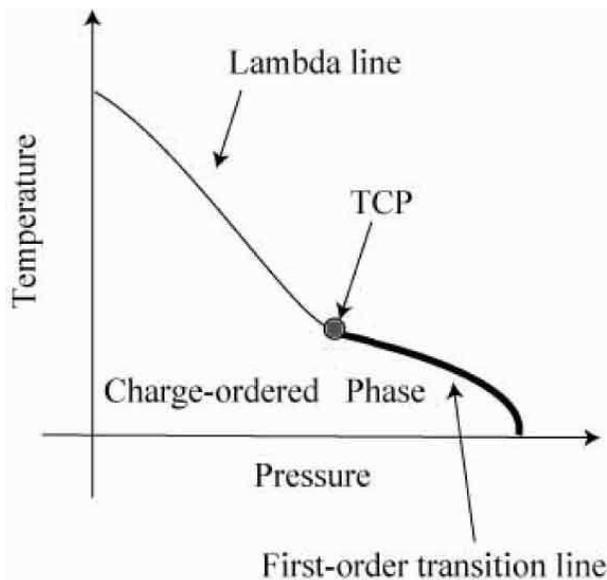}   
	\end{center}
\caption{Schematic phase diagram of (DI-DCNQI)$_2$Ag.}
\label{fig:Ex_Phase}
\end{figure}

Charge-order transitions take place either as first-order or continuous ones.
They even show both continuous and first-order boundaries meeting at a tricritical
point (TCP) in phase diagrams as in (DI-DCNQI)$_2$Ag~\cite{DCNQI} (see Fig. \ref{fig:Ex_Phase}). 
TCP is characterized as an endpoint of the first-order transition line. However, 
the first-order line does not simply terminate as the normal critical point because the charge-ordered 
phase explicitly breaks translational symmetry and the phase boundary should not disappear. 
After the termination of the first-order line, it continues as the continuous transition line
(the lambda line) and the meeting point of the first-order and the lambda lines is called TCP~\cite{TCP}. 
For (DI-DCNQI)$_2$Ag, the transition to the charge-ordered phase actually takes place with
a decrease in either temperature or pressure as we see in Fig. \ref{fig:Ex_Phase}.
This means that the charge-order transition may
be caused not only by suppressing thermal fluctuations but also as a quantum transition by 
suppressing the bandwidth with decreasing pressure. 
It offers an intriguing field of quantum critical phenomena for second-order and tricritical
transitions.
  
In this paper, we focus on various fluctuations expected around 
the continuous boundary (namely the lambda line) as well as around the TCP.
To the authors' knowledge, there exist no systematic studies on the question
which fluctuations and their divergences characterize the criticalities
of the charge-order transitions particularly around the TCP.
Divergences of fluctuations and singularities
of physical properties are the main subjects of this paper. 
We will show that for charge-order tricriticality, two completely different
fluctuations diverge: As is well known, one is the order parameter 
susceptibility at the ordering wavenumber, which leads to the Bragg 
scattering of the charge density response in the ordered phase.
The other divergence occurs in some of density fluctuations at vanishing wavenumber.  
This divergence also triggers the singularity of specific heat and further
nontrivial effects arise.
We will also show how the singularity of conductivity appears at the 
critical and tricritical points. At TCP, we will show that the singularity of conductivity
in the canonical ensemble is different from that in the grand-canonical ensemble.
Available experimental results on the conductivity are compared with the
results of present study.
The experimental results could be interpreted by the mean-field criticality of TCP
in the grand-canonical ensemble. 
We then propose further experiment to reveal the true tricriticality.
	
\begin{figure}[htp]
	\begin{center}
		\includegraphics[width=5cm]{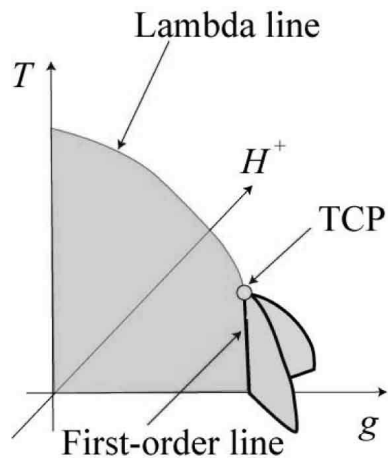}   
	\end{center}
\caption{Schematic phase diagram of TCP in the $T$-$g$-$H^{\dagger}$ space.
$T$, $g$ and $H^{\dagger}$ represent temperature, interaction and the field
which is conjugate to the order parameter, respectively. The shaded surfaces
represent the first-order transition surfaces.}
\label{fig:G_TCP}
\end{figure}

In the Ginzburg-Landau (GL) scheme, TCP is expressed by the $\phi^{6}$ theory~\cite{TCP}.
The free energy $F$ is expanded up to the sixth order with respect to the 
order parameter $m$ as 
\begin{equation}
	F=rm^{2}+um^{4}+vm^{6}.
\label{phi4}
\end{equation}
If $u>0$, $r=0$ represents a conventional
Ising-like critical point.
If $u<0$, three minimum states can emerge.
These three minima represent the coexistence of three phases.
Then three phase boundaries appear between each combination of phases
as first-order phase boundaries. Since the coexistence of two phases is
represented by a surface in the parameter space, the coexistence of three phases is represented by
a single curve, whose endpoint is TCP.
In other words, TCP appears as the crossing point of three lambda lines, where
the lambda lines are terminating lines of each first-order phase-boundary surface.
Figure \ref{fig:G_TCP} shows a schematic phase diagram of TCP in the $T$-$g$-$H^{\dagger}$ space,
where $T$, $g$ and $H^{\dagger}$ represent temperature, interaction and the field which is conjugate to
the order parameter, respectively.
Near TCP, $r$ and $u$ are described by the linear combination
of $|T-T_{c}|$ and $|g-g_{c}|$, where $T_{c}$ and $g_{c}$
represent the critical temperature of TCP and the critical interaction of TCP, respectively.
The first-order surfaces are depicted as shaded surfaces.  
In eq. (\ref{phi4}), $u=0$ and $r=0$ represent TCP.
TCP belongs to a different universality class 
from that of the continuous transition lines. 
It is also known that the upper critical dimension of TCP is three so that
fluctuations become irrelevant for $d>3$ and marginal at $d=3$. 
Therefore, the mean-field treatments are more or less justified in three dimensions except
for possible logarithmic corrections in the exponents. 

The GL free energy given in eq. (\ref{phi4}) describes the transition of the scalar
order parameter leading to Ising-type symmetry breaking.  The commensurate charge ordering 
occurring at a simple fractional filling of electrons is 
indeed represented by a discrete symmetry breaking of spatial translation.
For example, at quarter filling, the electron density is one per two unit cells and
the ground state is given by alternating electron-rich and electron-poor sublattice points with
double degeneracy.  If the spin degrees of freedom of electrons are ignored for the moment, 
the electron-rich and electron-poor sites are mapped to spin up and down Ising variables; 
they are mapped to the Ising model. 
Therefore, at least at the classical level without quantum fluctuations, 
this free energy (\ref{phi4})
provides a good starting point for general charge-order phenomena.

In quantum systems with itinerancy of electrons, however, a new aspect emerges, which is connected 
to the metal-insulator transition. In this paper, we study the first-order regime, 
where the metal-insulator transition occurs simultaneously with the charge-order transition
as will be illustrated later in Fig. \ref{fig:G_Phase}.
We also study the continuous transition, where the charge-order transition 
does not accompany the metal-insulator 
transition but only shows a crossover of transport properties. 
Then TCP appears also as the critical termination point of the metal-insulator transition.
We will study the singularity of the conductivity in all the regions in detail. 

In this paper, we only consider common features of the charge-order 
transition with a commensurate periodicity and do not discuss details that
depend on the detailed band structure, periodicity or charge-ordering pattern.
Criticality and tricriticality are well studied subjects of classical phase 
transitions~\cite{TCP},
while quantum effects particularly on the tricriticality are not well explored.
To study this issue using the microscopic model, 
we employ a simple 
extended Hubbard model defined by 
\begin{align}
H&=-\sum_{ij}t_{ij}(c_{i\sigma}^{\dagger}c_{j\sigma}+h.c.)
+U\sum_{i}n_{i\uparrow}n_{i\downarrow } \notag \\
&+V\sum_{\langle ij\rangle}N_{i}N_{j}
-\mu \sum_{i}N_{i},
\label{EHM}
\end{align}
which captures some essence and offers useful starting point of studies for charge-order transitions,
in general.
Here, $c_{i\sigma}^{\dagger} (c_{i\sigma})$ creates (annihilates) an electron with spin $\sigma$ at site $i$, 
respectively and 
$n_{i\sigma}$ and $N_{i}=(n_{i\uparrow}+n_{i\downarrow})$ are number operators.
The transfer $t_{ij}$ represents the electron hopping between $i$ and $j$ sites and the terms proportional to $U(V)$ represent
onsite (intersite) Coulomb interactions, respectively. For the intersite repulsion $V$,
we restrict ourselves to the repulsion of nearest-neighbor pairs $\langle i,j\rangle$
for $V$.

In the first step, we clarify phase diagram and criticality in the classical limit,
where we take $t_{ij}=0$. 
This classical model partially captures essential aspects of continuous charge-order transitions as well as TCP.
After confirmation of the mean-field behavior, we will show results of Monte Carlo studies
by using the exchange Monte Carlo algorithm~\cite{EXMC} to circumvent the critical slowing down. 
The critical exponents of various physical properties are studied. 

To consider quantum effects,
we next study the itinerant model by taking nonzero $t_{ij}$ by Hartree-Fock approximation and elucidate
the phase diagram in the plane of the temperature $T$, the interaction couplings $U$ and $V$ and 
the parameter to control the band structure $t_{ij}$.  We then clarify how the 
charge-order transition is coupled to the metal-insulator transition and how the 
criticality is described.

The physical property and susceptibility studied in this paper is summarized in the following:
\begin{description}
\item[1.] Order parameter is defined as 
\begin{equation}
m=\frac{1}{N_{s}}\sum_i N_ie^{i{\bf Q}{r}_i}
\end{equation}
with $N_{s}$ being number of the sites and ${\bf Q}=(\pi,\pi)$ for the square lattice, for example.
Order parameter susceptibility is defined as
\begin{equation} 
\chi_{m}=(\frac{d^{2}F}{dm^{2}})^{-1}.  
\end{equation}
The susceptibility of  $m^{2}$ is defined as
\begin{equation} 
\chi_{m^{2}}=(\frac{d^{2}F}{d(m^{2})^{2}})^{-1}.  \label{eq:chi_m2}
\end{equation}
\item[2.] Doublon density is defined as
\begin{equation}
\langle D({\bf 0})\rangle=\frac{1}{N_{s}}\frac{dF}{dU}=\frac{1}{N_{s}}\sum_i \frac{N_i(N_i-1)}{2} \label{eq:D}
\end{equation}
and the doublon susceptibility is defined as 
\begin{equation}
\chi_{D(\bf{0})}=-\frac{d^{2}F}{dU^{2}}. \label{eq:chi_D}
\end{equation}
\item[3.] The nearest-neighbor charge correlation
which is conjugate to the nearest-neighbor repulsion $V$
is defined as 
\begin{equation}
R=\frac{1}{N_{s}}\frac{dF}{dV}=\frac{1}{N_{s}}\sum_{<ij>}N_{i}N_{j}
\end{equation}
 and its susceptibility is defined as
\begin{equation}
\chi_{R}=\frac{d^{2}F}{dV^{2}}.
\end{equation}
\item[4.] Internal energy is defined as 
\begin{equation}
E=\frac{1}{N_{s}}\frac{d(F/T)}{d(1/T)}
\end{equation} 
and specific heat is defined as
\begin{equation}
C=\frac{dE}{dT}.
\end{equation} 
\item[5.] Charge density is defined as 
\begin{equation}
n=-\frac{1}{N_{s}}\frac{dF}{d\mu}=\frac{1}{N_{s}}\sum_{i}N_{i}
\end{equation}
and charge susceptibility is defined as 
\begin{equation}
\chi_{c}=-\frac{d^{2}F}{d\mu^{2}}.
\end{equation}
\end{description}

The Mean-field values of critical exponents  
on the lambda line  and TCP will be summarized
in the next section in Table \ref{Crex_MF}.

At TCP and on the lambda line, within the mean-field approximation,
we will show that $\langle D({\bf 0})\rangle$, $R$ and $n$ have
the same singularity as that of $m^{2}$.
Furthermore,  $\chi_{D{(\bf 0})}$, $\chi_{R}$ and $\chi_{c}$ diverge
with the same singularity as that of $\chi_{m^{2}}$ 
in general (see Table \ref{Crex}).
We will show that this relation still holds
in Monte Carlo calculations.
As shown below, only at half filling in the classical model, $n$ does not have the same singularity
as that of $m^{2}$  and
therefore  $\chi_{c}$ does not
diverge because of the particle-hole symmetry in both the mean-field approximation and Monte Carlo
calculations. 
Strong divergence of various fluctuations at zero wavenumber is an outstanding
feature of TCP in contrast to the lambda line.   

The organization of this paper is as follows: In \S \ref{Classical},
results of mean-field and Monte Carlo studies of the classical
model are shown. We specify  which types of fluctuations diverge
at TCP in the classical model. In \S \ref{Itinerant}, we clarify effects of itinerancy on TCP
and elucidate the criticality of the extended Hubbard model at the charge-order
transition. In \S \ref{Conductivity}, we analyze the singularity of
conductivity by Hartree-Fock approximation in detail by studying the singularity of the order 
parameter and compare the results with available experimental results. Section \ref{Summary} is devoted to summary and 
discussion.

\section{ Classical model}
\label{Classical}
\subsection{Model}
 First, we consider the classical limit of the extended Hubbard model. 	
In electronic systems, Coulomb repulsion plays a central role
in stabilizing the charge order. 
Therefore this classical model, which only considers Coulomb repulsion, 
captures important aspects of the charge order.
We take $t_{ij}=0$ in eq. (\ref{EHM}). Using the 
relation $n_{i\uparrow}n_{i\downarrow}=\left(N_{i}(N_{i}-1)\right)/2$, we obtain 
the classical limit of the extended Hubbard model:

\begin{equation}
	H_{cl}=U\sum_{i}\frac{N_{i}(N_{i}-1)}{2}+V\sum_{\langle ij\rangle}N_{i}N_{j}
-\mu\sum_{i}N_{i},
\label{eq:cl} 
\end{equation}
where $N_{i}$ takes 0, 1 or 2.

Here, we note that half filling $n=1$ is realized
by taking $\mu=zV+U/2$ from the symmetry of the Hamiltonian  (\ref{eq:cl}). 
Here, $z$ is the coordination number. To see this, we make the transformation 
 $N_i\rightarrow 2-N_i$. This mapping changes the chemical potential
$\mu$ into $\bar{\mu}=-\mu+U+2Vz$, while at half filling
this mapping does not change the total charge number. Therefore,
we obtain the relation $\bar{\mu}=\mu$. This leads to the relation
$\mu=zV+U/2$ for half filling.  

This model is equivalent to Blume-Emery-Griffith (BEG) model~\cite{BEG}. 
BEG model is used
for analyzing the tricritical behavior of $\rm{^{3}He}$-$\rm{^{4}He}$ mixture, where the
$\rm{^{3}He}$-$\rm{^{4}He}$ mixture is realized in three-dimensional space.
Since the upper critical dimension of TCP is three,
the critical exponents of TCP are correctly described by the mean-field treatment.
The mean-field treatment indeed succeeded in 
explaining the phase separation of the $\rm{^{3}He}$-$\rm{^{4}He}$ mixture
and the existence of TCP~\cite{BEG}.   

BEG model is defined as
\begin{equation}
	H_{\rm{BEG}}=-J\sum_{\langle ij\rangle}S_{i}S_{j}+\Delta\sum_{i}S_{i}^{2}-\mu^{\prime}\sum_{i}S_i,
\label{eq:BEG}
\end{equation}
where $S_{i}$ takes -1, 0 or 1.
The free energy of the BEG model is defined by 
\begin{equation}
	F_{\rm BEG}=-T\log Z_{BEG}=-T\log {\rm Tr}[e^{-H_{\rm BEG}/T}].
\end{equation}
Originally, the $\rm{^{3}He}$-$\rm{^{4}He}$ mixture is realized in continuum space, while BEG model is its
simplification to a lattice model, where $\rm{^{3}He}$ and $\rm{^{4}He}$ atoms occupy only discrete lattice points.
Here, the $S_i=0$ site corresponds to that occupied by $\rm{^{3}He}$ atoms and $S_i=\pm 1$ site corresponds 
to that occupied by $\rm{^{4}He}$ atoms.
Gauge degrees of freedom with $O(2)$ symmetry, which realizes superfludity of $\rm{^{4}He}$
by its symmetry breaking, is drastically simplified to two discrete degrees of freedom $S_{i}=\pm 1$.	

	Using the correspondence relations $N_{i}=S_{i}+1$, $U=\Delta$, $V=-J$ and  $U/2+2V-\mu=\mu^{\prime}$,
the classical extended Hubbard model (\ref{eq:cl}) and BEG model (\ref{eq:BEG}) are equivalent except
for a constant. 
	
\begin{figure}[htp]
	\begin{center}
		\includegraphics[width=8cm]{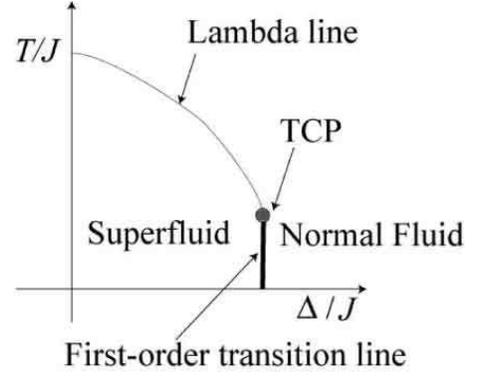}   
	\end{center}
\caption{Schematic phase diagram of BEG model. Thin line represents the continuous
transition line and thick line represents the first-order transition line.}
\label{fig:BEG_Phase}
\end{figure}

	In BEG model, the order parameter is $\langle S_{i}\rangle$. 
Here $\langle S_{i}\rangle\neq 0$ is interpreted as the superfludity of $\rm{^{4}He}$.
In large $\Delta$ region, where $\rm{^{3}He}$ concentration becomes higher,
the superfluid phase becomes less favored. 
Therefore, in the large $\Delta$ region, the normal state can coexist
with the superfluid state. Schematic phase diagram of normal and superfluid
in the $^{3}$He-$^{4}$He mixture is illustrated in Fig. \ref{fig:BEG_Phase}. 
Along the thick phase boundary, normal and superfluid phases coexist
which means  first-order transition across this phase boundary.
The thin curve describes continuous transition line, which meets
the first-order transition at the TCP. 

In BEG model, two different fluctuations diverge
at TCP. One is the fluctuation of the order parameter.
In fact, $\chi=-\partial^{2}F_{\rm{BEG}}/\partial {\mu^{\prime}}^{2}$ diverges
along the  continuous  transition and it also does at the TCP.
The other fluctuation does not diverge along the  continuous  transition line 
but does only at TCP. This fluctuation is represented by  
the susceptibility of $\langle S_{i}^{2}\rangle$, as defined by 
\begin{equation}
\chi_{2}=-\frac{\partial^{2}F_{\rm{BEG}}}{\partial \Delta^{2}}.
\end{equation}

The divergence of two fluctuations is an outstanding feature of TCP.
In our classical model, $\Delta$ corresponds to
$U$. Therefore the latter fluctuation corresponds to that of the doublon density
$\langle D({\bf0})\rangle=\langle n_{i\uparrow}n_{i\downarrow}\rangle$, 
because $n_{i\uparrow}n_{i\downarrow}$ is the conjugate quantity to $U$ in (\ref{EHM}).
The divergence of the doublon susceptibility $\chi_{D({\bf0})}$ defined in eq. (\ref{eq:chi_D})
is in fact analyzed in the next section within the mean-field approximation.

\subsection{Mean-field approximation}
\label{sec:MF}
In this section, we show results of the mean-field approximation
just for comparison with Monte Carlo results and results 
on the itinerant model in later sections.
The mean-field approximation reproduces the Ginzburg-Landau approximation of the 
$\phi^{6}$ theory. In the mean-field approximation,
we decouple the interaction term in eq. (\ref{eq:cl}) as
\begin{align}
	N_{i}N_{j}&=(N_{i}-\langle N_{i}\rangle+\langle N_{i}\rangle)(N_{j}-\langle N_{j}\rangle+\langle N_{j}\rangle) \notag \\
					&\sim N_{i}\langle N_{j}\rangle+N_{j}\langle N_{i}\rangle-\langle N_{i}\rangle\langle N_{j}\rangle. 
\end{align}

 We assume $\langle N_{i}\rangle=n\pm m$. The upper sign $+$ is for
the A sublattice and the lower sign  $-$ is for the B sublattice on a bipartite lattice.
Using this mean field, we obtain the mean-field Hamiltonian
\begin{align}
	H_{MF}&=\frac{U}{2}\sum_{i}N_{i}(N_{i}-1)+Vz\sum_{i \in A}N_{i}(n+m) \notag \\
&+Vz\sum_{i \in B}N_{i}(n-m)-\mu\sum_{i}N_{i}-\frac{N_{s}Vz(n^2-m^2)}{2},
\label{eq:H_MF}
\end{align}
where $z$ is the coordination number.
From eq. (\ref{eq:H_MF}), we obtain the free energy per site as
\begin{align}
 F_{MF}&=-TN_{s}Z_{MF} \notag \\
				&=-\frac{T}{2 N_{s}} 
			\log(1+a(m)+b(m)) \notag \\
			&-\frac{T}{2 N_{s}} 
			\log(1+a(-m)+b(-m)) \notag \\
			&-\frac{V^{\prime}}{2}(n^2-m^2),
\end{align}
where  $V^{\prime}=Vz$ and $a(m)$ and $b(m)$ are defined as
\begin{align}
	a(m)&=e^{-(V^{\prime}(n+m)-\mu)/T}, \\ 
	b(m)&=e^{-(U+2V^{\prime}(n+m)-2\mu)/T}. 
\end{align}

 The minimum condition $dF_{MF}/dm=0$ leads to the self-consistent equation
\begin{equation}
 m=\frac{1}{2}(\frac{a(-m)+2b(-m)}{1+a(-m)+b(-m)}-\frac{a(m)+2b(m)}{1+a(m)+b(m)}).
\end{equation} 
 
	From this free energy, the charge density $n$ and the doublon density $\langle D(\bf{0})\rangle$
are respectively given as
\begin{align}
	n&=-\frac{\partial F_{MF}}{\partial \mu} \notag \\
	&=\frac{1}{2}(\frac{a(m)+2b(m)}{1+a(m)+b(m)}+\frac{a(-m)+2b(-m)}{1+a(-m)+b(-m)}),
\label{eq:n}
\end{align}
\begin{align}
	\langle D(\bf{0})\rangle&=\frac{\partial F_{MF}}{\partial U} \notag \\
	&=\frac{1}{2}(\frac{a(m)}{1+a(m)+b(m)}+\frac{a(-m)}{1+a(-m)+b(-m)}).  
\label{eq:MF_D}
\end{align}

 We also obtain the internal energy $E$ and the next nearest charge correlation $R$ which is
conjugate to $V$ as  
\begin{align}
	E&=\frac{\partial (F_{MF}/T)}{\partial(1/T)} \notag \\
	&=V^{\prime}(n^{2}+m^{2})+\frac{U}{2}(n-\langle D({\bf0})\rangle)-\mu n, \label{eq:E} \\  
	R&=\frac{\partial F_{MF}}{\partial V} \notag \\
	&=\sum_{<ij>}\langle N_{i}\rangle\langle N_{j}\rangle \notag \\ 
	&=\frac{z}{2}(n^{2}-m^{2}). \label{eq:R}  
\end{align}

 For the moment, we consider half-filled case, $n=1$. 
As mentioned above, half filling is realized by taking  
$\mu=V^{\prime}+U/2$  independent of $T$ or $m$.
Therefore, we can easily expand the free energy with respect
to the order parameter $m$ at half filling. By taking $n=1$ and
$\mu=V^{\prime}+U/2$, GL expansion is reproduced as
\begin{equation}
	F=A+rm^{2}+um^{4}+vm^{6}-mH^{\dagger},
\label{eq:F_MF}
\end{equation}
where $A$, $r$, $u$ and $v$ are respectively defined as
\begin{align}
	A&=-V^{\prime}/2-\log(2+e^{U/2T}) \\ 
	r&=\frac{V^{\prime}(2+e^{U/2T}-2V^{\prime}/T)}{2(2+e^{U/2T})}  \label{eq:r}\\ 
	u&=\frac{{V^{\prime}}^{4}(4-e^{U/2T})}{12T^{3}(2+e^{U/2T})^{2}} \label{eq:u} \\
	v&=\frac{{V^{\prime}}^{6}(-64+26e^{U/2T}-e^{U/T})}{360T^{5}(2+e^{U/2T})^{3}},
\end{align} 
and $H^{\dagger}$ is the field conjugate to $m$.

{\bf Lambda line: half filling}

Now, we consider the criticality of physical properties along the lambda line at half filling.
As mentioned above, the lambda line is represented by the condition of $r=0$ and $u>0$.
Along the lambda line, the criticality of the order parameter and the susceptibility of 
the order parameter $\chi_{m}$ are defined by
\begin{align}
	m&\sim r^{\beta}, \\
	\chi_{m}&\sim r^{-\gamma}.
\end{align}
Since $m=\partial F_{MF}/\partial H^{\dagger}$ and $\chi_{m}=(\partial^{2} F_{MF}/\partial m^{2})^{-1}_{H^{\dagger}=0}$,
we obtain from eq. (\ref{eq:F_MF})  
\begin{align}
	&\beta=\frac{1}{2} \label{eq:m_lam} \\
	&\gamma=1. 
\end{align}  

Along the lambda line, the criticality of $\langle D({\bf 0})\rangle$ in eq. (\ref{eq:D})
and $\chi_{D({\bf 0})}$ in eq. (\ref{eq:chi_D}) are defined by
\begin{align}
	\langle D({\bf 0})\rangle&\sim r^{\beta_{2}} \\ 
	\chi_{D({\bf 0})}&\sim r^{-\gamma_{2}}. 
\end{align} 
 
 In the mean-field scheme, the doublon density $\langle D(\bf{0})\rangle$ is proportional
to $m^{2}$ except for a constant in the critical region. From eq. (\ref{eq:MF_D}), 
we obtain the expansion of $\langle D(\bf{0})\rangle$ with respect to $m$ as
\begin{equation}
  \langle D({\bf 0})\rangle\sim a_{0}+a_{1}m^{2},
\end{equation}
where $a_{0}$ and $a_{1}$ are defined by
\begin{align}
 a_{0}&=\frac{1}{2+e^{-U/2T}} \\
 a_{1}&=\frac{{V^{\prime}}^{2}e^{U/2T}}{2T^{2}(1+2e^{-U/2T})^{2}}.
\end{align}

Therefore, the singularity of the doublon density
is equivalent to that of  $m^{2}$ and its susceptibility $\chi_{D(\bf{0})}$ is equivalent to
$\chi_{m^{2}}$.
From eqs. (\ref{eq:m_lam}) and (\ref{eq:chi_m2}), along the lambda line, 
critical exponents of the doublon density $\langle D(\bf{0})\rangle$
and its susceptibility $\chi_{D(\bf{0})}$ are obtained as 
\begin{align}
	\beta_{2}=1 \\ 
	\gamma_{2}=0, \label{eq:MF_gamma_2} 
\end{align} 
where eq. (\ref{eq:MF_gamma_2}) indicates that $\chi_{ D(\bf{0})}$ does not diverge along the lambda line.

Equations (\ref{eq:E}) and (\ref{eq:R}) show that $E$ and $R$ have the same singularity as that 
of $m^2$ whereas the specific heat $C$ and the susceptibility of $R$($\chi_{R}$) have the same singularity
as that of $\chi_{m^{2}}$ for the density fixed at $n=1$. 
Therefore, along the lambda line, $C$ and $\chi_{R}$ does not
diverge.  

\begin{figure}[htp]
	\begin{center}
		\includegraphics[width=6cm]{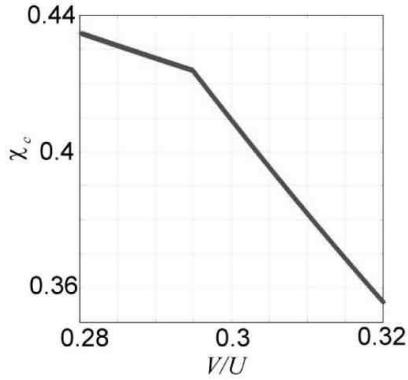}   
	\end{center}
\caption{Charge susceptibility $\chi_{c}$ is plotted  near the lambda line at half filling.
We fix temperature at $T/U=0.5$.}
\label{fig:Half_lam_chi_C}
\end{figure}

Here, we consider the criticality of the charge susceptibility $\chi_{c}$ at the lambda transition.
Equation (\ref{eq:F_MF}) expresses the expansion when the density is fixed
at $n=1$, while $\chi_{c}$ is obtained by changing the density
infinitesimally  from $n=1$.
Therefore, the singularity of $\chi_{c}$ cannot
straightforwardly be obtained from eq. (\ref{eq:F_MF}). 
To clarify the singularity of $\chi_{c}$,
we calculate $\chi_{c}$ numerically near
the lambda line as shown in Fig. \ref{fig:Half_lam_chi_C}.
We confirm that $\chi_{c}$ does not diverge at the lambda transition within the mean-field approximation.  

{\bf TCP: half filling}

Next, we consider the criticality of physical properties near TCP at half filling.
TCP is represented by the condition of $r=0, u=0$ in eq. (\ref{eq:F_MF}).
From this condition, the location of TCP is determined from eqs. (\ref{eq:r}) and (\ref{eq:u}) 
as $T_{c}=1/(4\log2)$, $V_{c}=3/(16\log2)$. 
At TCP, the critical exponents of the order parameter and the susceptibility of the 
order parameter $\chi_{m}$  at TCP are defined by 
\begin{align}
	m&\sim r^{\beta_t} \\  
	\chi_{m}&\sim r^{-\gamma_t}. 	
\end{align}
Taking $u=0$, we obtain from eq. (\ref{eq:F_MF}) as 
\begin{align}
	&\beta_{t}=\frac{1}{4}  \label{eq:m_TCP} \\ 
	&\gamma_{t}=1. 
\end{align}  

The critical exponents
of $\langle D(\bf{0})\rangle$ and $\chi_{D(\bf{0})}$ at TCP  are defined by
\begin{align}
	\langle D(\bf{0})\rangle&\sim r^{\beta_{2t}} \\  
	\chi_{D(\bf{0})}&\sim r^{-\gamma_{2t}}. 	
\end{align}
Again we note that, within the mean-field approximation,
the singularity of $\langle D(\bf{0})\rangle$ and $\chi_{D(\bf{0})}$ are also equivalent to that of
$m^{2}$ and $\chi_{m^{2}}$ near TCP.
Using $\beta_{t}=\frac{1}{4}$ and the definition of $\chi_{m^{2}}$
in eq. (\ref{eq:chi_m2}),
we obtain the critical exponents as
\begin{align}
	\beta_{2t}&=2\beta_{t}=\frac{1}{2} \\
	\gamma_{2t}&=\frac{1}{2}.
\end{align}

From eqs. (\ref{eq:E}) and (\ref{eq:R}),
the internal energy $E$ and $R$ have the same singularity
as that of $m^{2}$ and their susceptibility have the same singularity as that of $\chi_{m^{2}}$.
From eq. (\ref{eq:m_TCP}), at TCP, we obtain the singularity of $\langle D(\bf{0})\rangle$, $E$ and $R$ as
\begin{align}
 \langle D({\bf0})\rangle, E, R\sim m^{2}\sim r^{\beta_{2t}},\label{eq:D_TCP} \\
 \chi_{\langle D(\bf{0})\rangle}, C, \chi_{R}\sim \chi_{m^{2}}\sim r^{-\gamma_{2t}},\label{eq:chi_D_TCP} 
\end{align}
Equations (\ref{eq:D_TCP}) and (\ref{eq:chi_D_TCP})
lead to $\beta_{2t}=1/2$ and $\gamma_{2t}=1/2$.
Because $\gamma_{2t}$ is positive, $\chi_{D(\bf{0})}$, $C$ and $\chi_{R}$ diverge at TCP
in contrast to the lambda transition.

{\bf Charge susceptibility at TCP}

\begin{figure}[htp]
	\begin{center}
		\includegraphics[width=6cm]{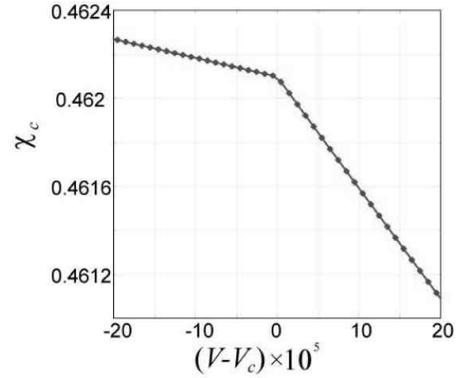}   
	\end{center}
\caption{Charge susceptibility $\chi_{c}$ is plotted near TCP at half filling
within the mean-field approximation.
 We fix temperature at $T_{c}=1/(4\log 2)$. 
Critical interaction is given by $V_{c}=3/(16\log 2)$.}
\label{fig:Half_chi_C}
\end{figure}

 We now consider the charge susceptibility $\chi_{c}$ at TCP within the mean-field
scheme. The charge susceptibility is defined by $\chi_{c}=-\partial^{2}F_{MF}/\partial \mu^{2}$.
 We confirm that $\chi_{c}$ at half filling does not diverge at TCP 
because of the particle-hole symmetry (see section \ref{chi_c}).
Charge susceptibility $\chi_c$ at half filling is shown in Fig. \ref{fig:Half_chi_C}.

\begin{figure}[htp]
	\begin{center}
		\includegraphics[width=6cm]{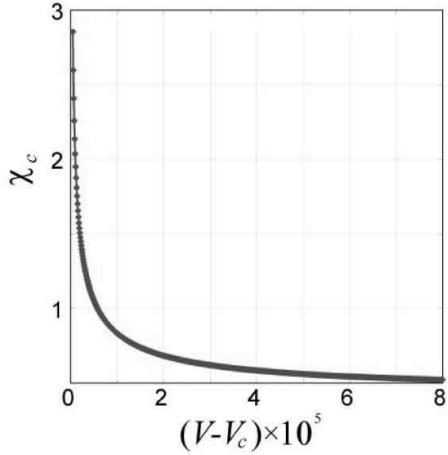}   
	\end{center}
\caption{Charge susceptibility  $\chi_{c}$ is plotted near TCP at $\mu/U=1.2$
within the mean-field approximation. Here, $T_{c}/U$ and $V_{c}/U$
are estimated as $0.31356$ and $0.2663996$, respectively.}
\label{fig:MF_chi_C}
\end{figure}

\begin{figure}[htp]
	\begin{center}
		\includegraphics[width=8cm]{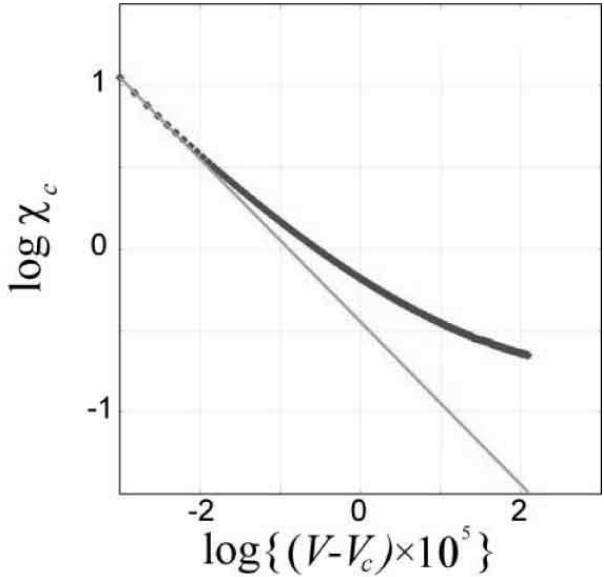}   
	\end{center}
\caption{Scaling plot of $\chi_{c}$  near TCP within the mean-field approximation.
Slope determines the exponent for $\chi_{c}$.
We estimate $\chi_{c}=|V-V_{c}|^{-0.499(1)}$.
This value is consistent with
the mean-field value $\gamma_{2t}=1/2$.}
\label{fig:MF_chi_C_scale}
\end{figure}

 Next, we consider the TCP away from half filling.
If we take $\mu/U=1.2$, we confirm that $\chi_{c}$ diverges at TCP.
Figures \ref{fig:MF_chi_C} and \ref{fig:MF_chi_C_scale} show $\chi_c$ at TCP.
We estimate the criticality of $\chi_{c}$ as $\chi_{c}\sim |V-V_{C}|^{-0.499(1)}$.
This value is consistent with the mean-field value $\gamma_{2t}=1/2$.
This result indicates that the singularity of the charge density $n$
coincides with that of $m^{2}$.
This implies that the free energy can be expanded as
\begin{equation}
F\sim A^{\prime}-\mu^{\prime} n+u^{\prime}n^{2}+v^{\prime}n^{3}+\dotsm
\end{equation}
as a functional of $n$ with 
$A^{\prime}$, $\mu^{\prime}$, $u^{\prime}$ and $v^{\prime}$
being constants.
Therefore, $\chi_{c}$ has the same singularity
as that of $\chi_{m^{2}}$ at TCP, in general.  

Here, in Table \ref{Crex_MF}, we summarize the exponents of physical properties
obtained by the mean-field approximation.
\begin{table}
\begin{center}
	\begin{tabular}{|l|c|c|} \hline
		\multicolumn{1}{|c|}{ }  &\text{Lambda line} & \text{TCP} \\ \hline
		\multicolumn{1}{|c|}{\text{$m$}}  & \text{$\beta=\frac{1}{2}$} & \text{$\beta_{t}=\frac{1}{4}$} \\ \hline
		\multicolumn{1}{|c|}{\text{$\langle D({\bf0})\rangle\sim m^{2}$}} 
		& \text{$\beta_{2}=2\beta=1$}& \text{$\beta_{2t}=2\beta_{t}=\frac{1}{2}$}  \\ 		\hline		  
		\multicolumn{1}{|c|}{\text{$E$}}  & \text{$1$} & \text{$\frac{1}{2}$} \\ \hline
		\multicolumn{1}{|c|}{\text{$R$}}  & \text{$1$} & \text{$\frac{1}{2}$} \\ \hline
		\multicolumn{1}{|c|}{\text{$n$}}  & \text{constant} & \text{constant} \\ \hline
		\multicolumn{1}{|c|}{\text{$\chi_{m}$}} 
		& \text{$\gamma=1$} & \text{$\gamma_{t}=1$} \\ \hline
		\multicolumn{1}{|c|}{\text{$C$}}  & \text{$\alpha=0$} & \text{$\alpha_{t}=\gamma_{2t}=\frac{1}{2}$} \\ \hline
		\multicolumn{1}{|c|}{\text{$\chi_{D(\bf{0})}\sim\chi_{m^{2}}$}} 
		 & \text{$\gamma_{2}=0$}& \text{$\gamma_{2t}=\frac{1}{2}$}  \\ \hline
		\multicolumn{1}{|c|}{\text{$\chi_{R}$}}  & \text{$0$} & \text{$\frac{1}{2}$} \\ \hline
		\multicolumn{1}{|c|}{\text{$\chi_{c}$}}  & \text{cusp} & \text{cusp*} \\ \hline
	\end{tabular}
\end{center}
\caption{Mean-field values of physical properties at half filling  on the lambda line and at TCP.
*Away from half filling ($\mu/U=1.2$), 
$\chi_{c}$ diverges with the singularity $\chi_{c}\sim |V-V_{c}|^{-0.499\pm 0.001}$.}
\label{Crex_MF}
\end{table}%

\subsection{Monte Carlo results: Phase diagram}
 To study this classical model beyond the mean-field approximation,
we perform exchange Monte Carlo~\cite{EXMC} calculations
on the two-dimensional square lattice. We note that the Monte Carlo
results give exact results within statistical errors.

 The exchange Monte Carlo calculation is performed in the following way:
First, we prepare replicas that have different temperatures.  
On each replica, we perform a standard Monte Carlo simulation and 
renew each state. Then, we exchange the configurations of each replica 
by the extended probability distribution function $W$, which is defined as 
\begin{align}
 &\Delta(X_n,X_m;T_m,T_n)=(1/T_n-1/T_m)\left(E(X_n)-E(X_m)\right), \\ 
 &W(X_n,X_m;T_m,T_n)=\frac{1}{1+\exp(\Delta)},
\end{align}
where $X_n$, $E(X_n)$, $T_n$ represent the configuration of the $n$th-replica,
the internal energy of the $n$th-replica, and the temperature of the $n$th-replica, respectively. 
This extended probability distribution function $W$
satisfies the detailed balance condition 
\begin{equation}
 \frac{W(X_n,X_m;T_m,T_n)}{W(X_n,X_m;T_n,T_m)}=\exp(-\Delta).
\end{equation}
This method is efficient in overcoming the critical slowing down.  

 At TCP, the upper critical dimension is at $d_{c}=3$.
Therefore, the mean-field approach is justified in three dimensions
except for logarithmic corrections. 
On the other hand, in 2D, the mean field theory is not accurate anymore.
Here, we perform Monte Carlo calculation in 2 dimensions
on a square lattice.  
 
\begin{figure}[htp]
	\begin{center}
		\includegraphics[width=8cm]{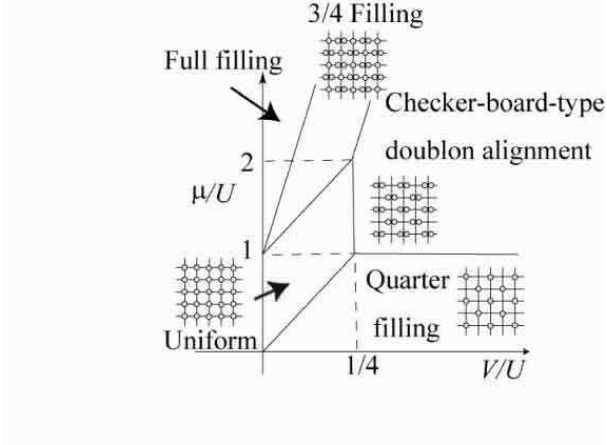}   
	\end{center}
\caption{Ground state phase diagram of the classical model.
``Full filling'' means a phase where all the sites are doubly occupied by particles.
``Uniform'' means a phase where each site is occupied by one particle.
``3/4 filling'' represents alternating singly and doubly occupied sites in a staggered pattern.
Phase boundaries between full filling and 3/4 filling, 3/4 filling and uniform, 
uniform and quarter filling are given as $\mu/U=8V/U+1$, $\mu/U=4V/U+1$, $\mu/U=4V/U$, respectively.}
\label{fig:Ground}
\end{figure}

	Ground-state phase diagram is shown in Fig. \ref{fig:Ground}.
It should be noted that various charge-order states
appear even in the classical model that considers only 
the onsite repulsion $U$ and the nearest-neighbor repulsion $V$.

\begin{figure}[htp]
	\begin{center}
		\includegraphics[width=7cm]{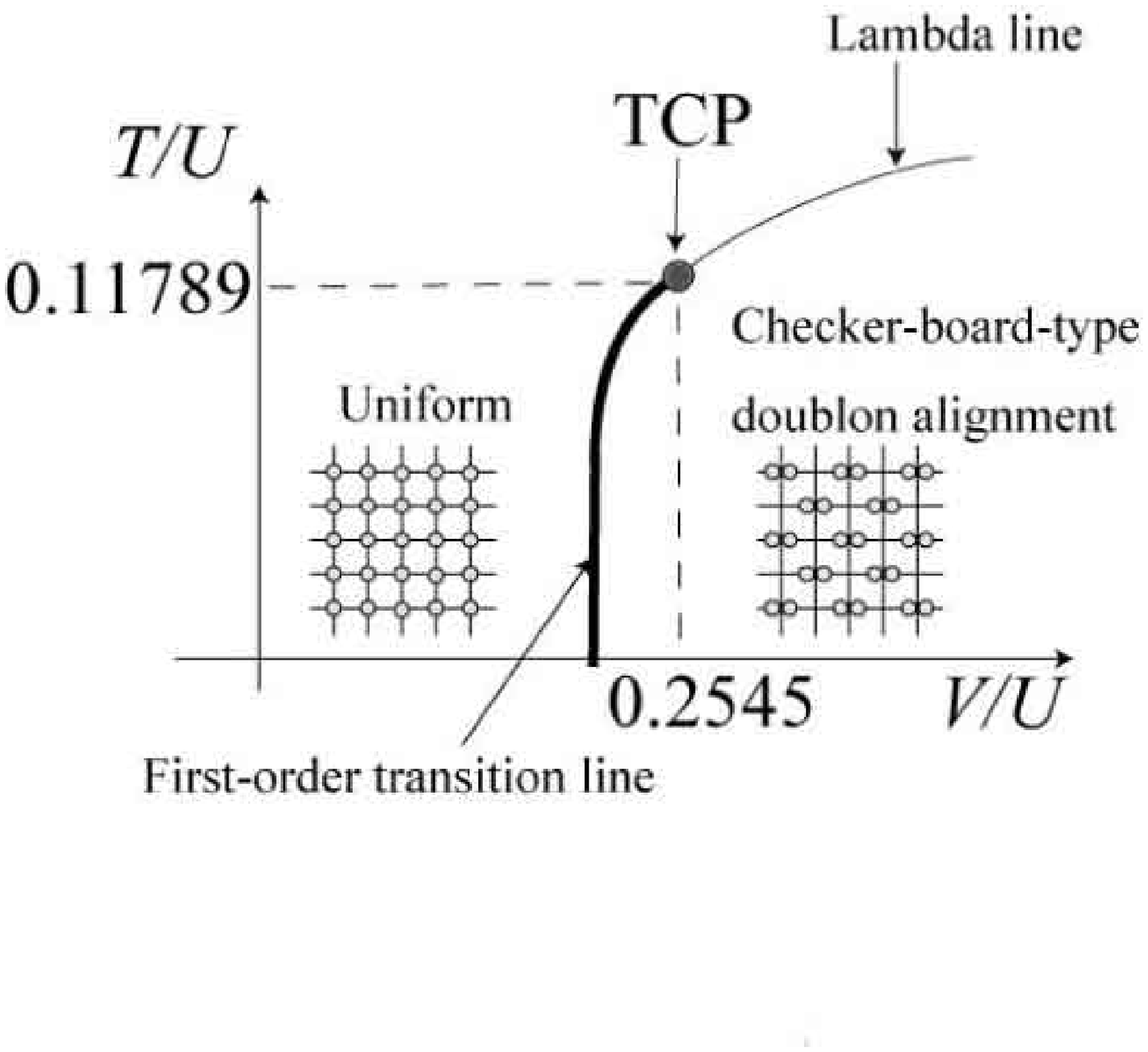}   
	\end{center}
\caption{Phase diagram in the parameter space of scaled
temperature $T/U$ and the intersite interaction $V/U$
at $\mu/U=1.2$.}
\label{fig:Phase}
\end{figure}

At finite temperatures, we find TCP as in Fig. \ref{fig:Phase}
at the phase boundary between the charge order with 
checker-board-type doublon alignment and the uniform phase.
The order parameter of this charge order with checker-board-type doublon alignment at the wavenumber
$\bf{Q}=(\pi,\pi)$ is given by
\begin{equation}
	m=\sum_{i}N_i e^{i\mathbf{Qr_{i}}}. 
\end{equation} 

\subsection{Monte Carlo results: Singularities and exponents}
We consider the case $\mu/U=1.2$ as an example of a generic case. 
As shown below, $\chi_{c}$ at half filling has a different singularity
from that at general fillings,
while the singularities of other fluctuations are independent of fillings.
Therefore, the case at $\mu/U=1.2$ captures the general nature of TCP.  
We will give the reason for the nontrivial behavior of $\chi_{c}$ in detail later. 
   
First, we obtain the critical exponents $\nu$ and $\eta$ 
for finite-size scalings defined by
\begin{align}
	&M=\langle m^{2}\rangle= L^{2-\eta}\Phi_{1}(tL^{1/\nu}), \label{eq:eta}\\	
 	&A=\frac{d\ln\langle m^{2}\rangle}{d(1/T)}= L^{\frac{1}{\nu}}\Phi_{2}(tL^{1/\nu}).\label{eq:nu}
\end{align}
Here, $L$ is the linear dimension of the 
lattice size, $\tilde{t}$ is the reduced temperature 
defined by $\tilde{t}=(T-T_{c})/T_{c}$ and $\Phi_{1}$ and $\Phi_{2}$
are scaling functions.  
As shown in Table \ref{HS_Lambda} and \ref{HS_TCP}, 
using hyperscaling relations~\cite{TCP},
other exponents defined in Table \ref{Crex_MF}
are described by using $\nu$ and $\eta$.

\begin{table}
\begin{center}
	\begin{tabular}{|l|c|c|c|} \hline
		\multicolumn{1}{|c|}{\text{Exponent} }  & \text{Hyperscaling relation} 
		& \text{2D Ising value} 	\\ \hline
		\multicolumn{1}{|c|}{\text{$\nu$}} 
		 & \text{} & \text{$1$}  \\ \hline
		\multicolumn{1}{|c|}{\text{$\eta$}} 
		 & \text{$$} & \text{$\frac{1}{4}$}  \\ \hline
		\multicolumn{1}{|c|}{\text{$\alpha$}} 
		 & \text{$2-d\nu$} & \text{$0$}  \\ \hline
		\multicolumn{1}{|c|}{\text{$\beta$}} 
		  & \text{$\frac{\nu(d-2+\eta)}{2}$}& \text{$\frac{1}{8}$}  \\ \hline
		\multicolumn{1}{|c|}{\text{$\gamma$}} 
		  & \text{$(2-\eta)\nu$} & \text{$\frac{7}{4}$}  \\ \hline
	\end{tabular}
\end{center}
\caption{Hyperscaling relations on the lambda line.
$d$ represents spatial dimension. 
Exponents of 2D Ising model are obtained from exact solution~\cite{Kaufmann}.}
\label{HS_Lambda}
\end{table}%

\begin{table}
\begin{center}
	\begin{tabular}{|l|c|c|} \hline
		\multicolumn{1}{|c|}{\text{Exponents} }  & \text{Hyperscaling relations} 
		& \text{2D BEG model values} 	\\ \hline
		\multicolumn{1}{|c|}{\text{$\nu_{t}$}} 
		 & \text{} & \text{$0.56(1)$}  \\ \hline
		\multicolumn{1}{|c|}{\text{$\eta_{t}$}} 
		 & \text{$$} & \text{$0.14(2)$}  \\ \hline
		\multicolumn{1}{|c|}{\text{$\alpha_{t}$}} 
		  & \text{$2-d\nu_{t}$} & \text{$0.89(2)$}  \\ \hline
		\multicolumn{1}{|c|}{\text{$\beta_{t}$}} 
		  & \text{$\frac{\nu_{t}(d-2+\eta_{t})}{2}$} & \text{$0.039(6)$}  \\ \hline
		\multicolumn{1}{|c|}{\text{$\beta_{2t}$}} 
		  & \text{$\nu_{t}d-1$} & \text{$0.12(2)$} \\ \hline
		\multicolumn{1}{|c|}{\text{$\gamma_{t}$}} 
		  & \text{$(2-\eta_{t})\nu_{t}$} & \text{$1.03(2)$} \\ \hline
		\multicolumn{1}{|c|}{\text{$\gamma_{2t}$}} 
		 & \text{$2-d\nu_{t}=\alpha_{t}$} & \text{$0.89(2)$}  \\ \hline
	\end{tabular}
\end{center}
\caption{Hyperscaling relations at TCP~\cite{TCP}.
$d$ represents spatial dimension.
Exponents of 2D BEG model are obtained numerically~\cite{MCRG}.
The parentheses
denote the error bars in the last digit.}
\label{HS_TCP}
\end{table}%

\subsubsection{Exponents $\nu$ and $\eta$ analyzed from finite-size scaling}
 
{\bf Lambda line: $\mu/U=1.2$}

 On the lambda line, the universality class of this phase transition is
categorized to that of the conventional 2D Ising model~\cite{Nigel}.
As described below, we indeed confirmed that critical exponents
are the same as that of the 2D Ising model derived from 
Onsager's exact solution~\cite{Kaufmann}. 

To obtain critical exponents, we use finite-size scaling.
From eqs. (\ref{eq:eta}) and (\ref{eq:nu}), at $\tilde{t}=0$, 
$\log[M(L_{1})/M(L_{2})]/\log[L_1/L_2]$
has the universal value $2-\eta$, which is independent of lattice size.
Similarly, $\log[A(L_1)/A(L_2)]/\log[L_1/L_2]$ converges to its universal value $1/\nu$. 

\begin{figure}[htp]
	\begin{center}
		\includegraphics[width=6cm]{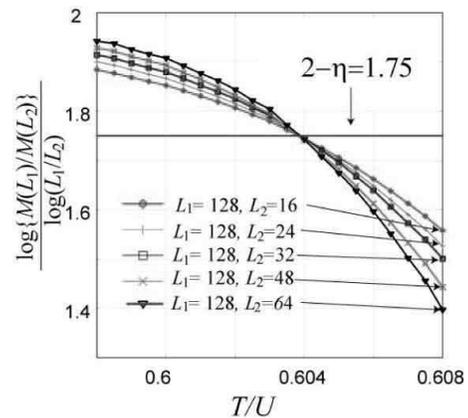}   
	\end{center}
\caption{Result of finite-size scaling near the lambda line.
The parameters are  $L_{1}=128, L_{2}=16$ (open diamonds),
 $L_{1}=128, L_{2}=24$($+$),
 $L_{1}=128, L_{2}=32$ (squares),
$L_{1}=128, L_{2}=48$($\times$) and
$L_{1}=128, L_{2}=64$ (triangles).
The horizontal line  shows the exact critical exponent.}
\label{fig:lambda_eta}
\end{figure}

\begin{figure}[htp]
	\begin{center}
		\includegraphics[width=7cm]{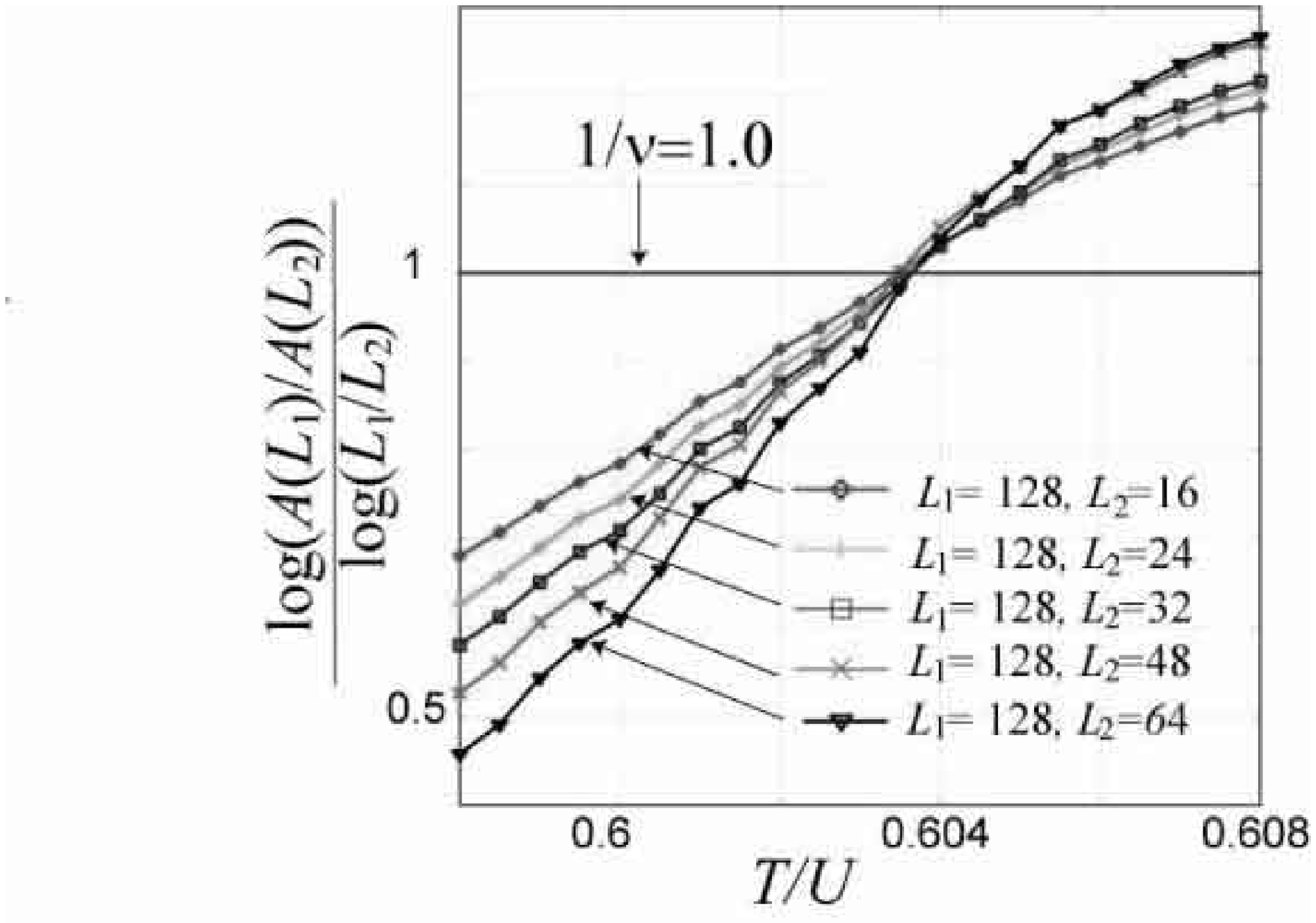}   
	\end{center}
\caption{Result of finite-size scaling near the lambda line.
The parameters are  $L_{1}=128, L_{2}=16$ (open diamonds),
 $L_{1}=128, L_{2}=24$($+$),
 $L_{1}=128, L_{2}=32$ (squares),
$L_{1}=128, L_{2}=48$($\times$) and
$L_{1}=128, L_{2}=64$ (triangles).
The horizontal line  shows the exact critical exponent.}
\label{fig:lambda_nu}
\end{figure}

The result of finite-size scaling at $V/U=0.6$ is 
shown in Figs. \ref{fig:lambda_eta} and  \ref{fig:lambda_nu}.
We obtain $\eta=0.25(1)$ and $\nu=1.03(5)$.
We estimate the error by the standard deviation of the crossing points.
These results are consistent with the exact values $\eta=1/4$ and $\nu=1$. 

{\bf TCP: $\mu/U=1.2$}

\begin{figure}[htp]
	\begin{center}
		\includegraphics[width=7cm]{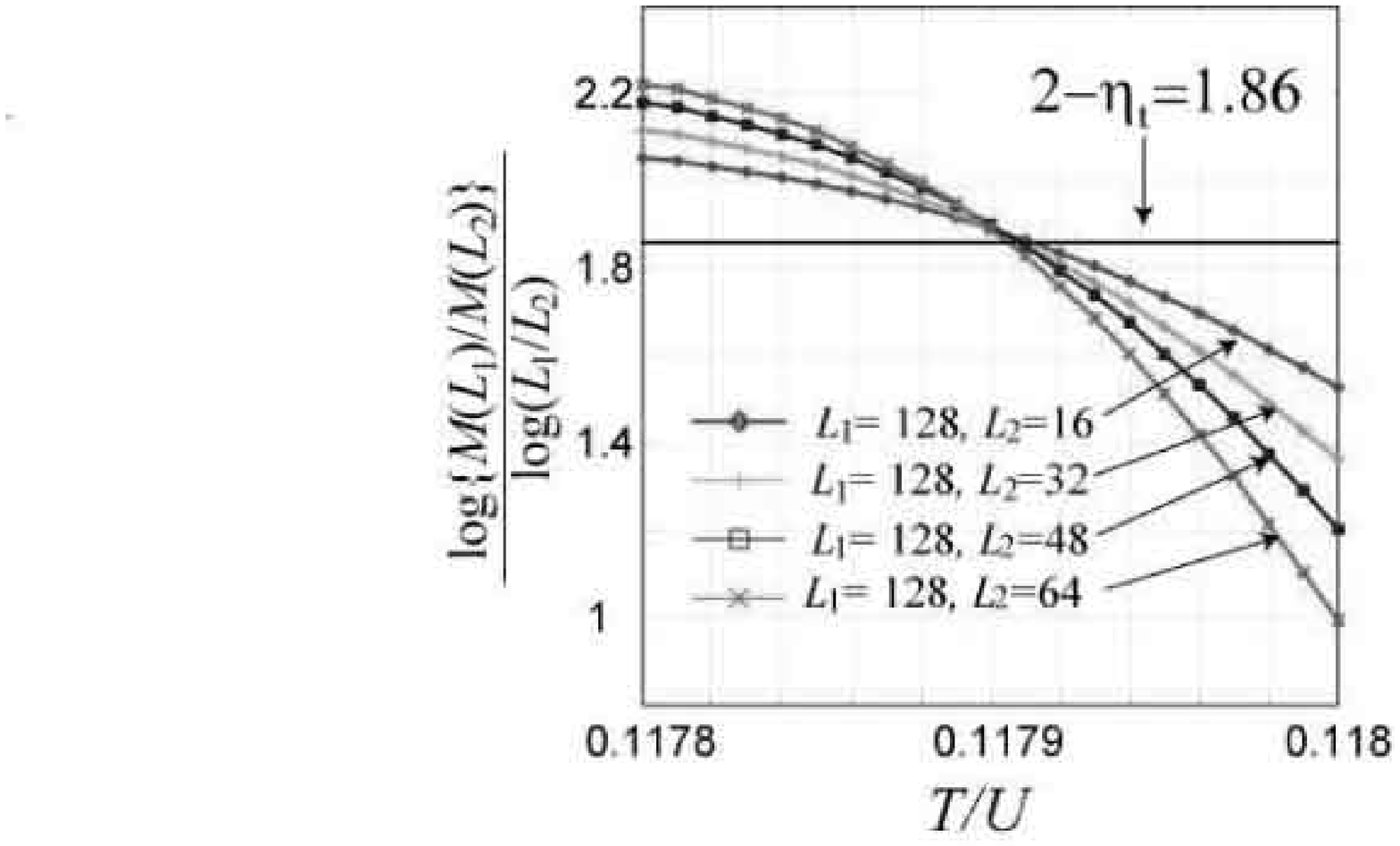}   
	\end{center}
\caption{Finite-size scaling of $M$ near TCP.
The parameters are  $L_{1}=128, L_{2}=16$ (open diamonds),
 $L_{1}=128, L_{2}=32$($+$),
 $L_{1}=128, L_{2}=48$ (squares),
$L_{1}=128, L_{2}=64$($\times$).
The horizontal line  shows the critical exponent in Ref. 10.}
\label{fig:TCP_eta}
\end{figure}

\begin{figure}[htp]
	\begin{center}
		\includegraphics[width=6cm]{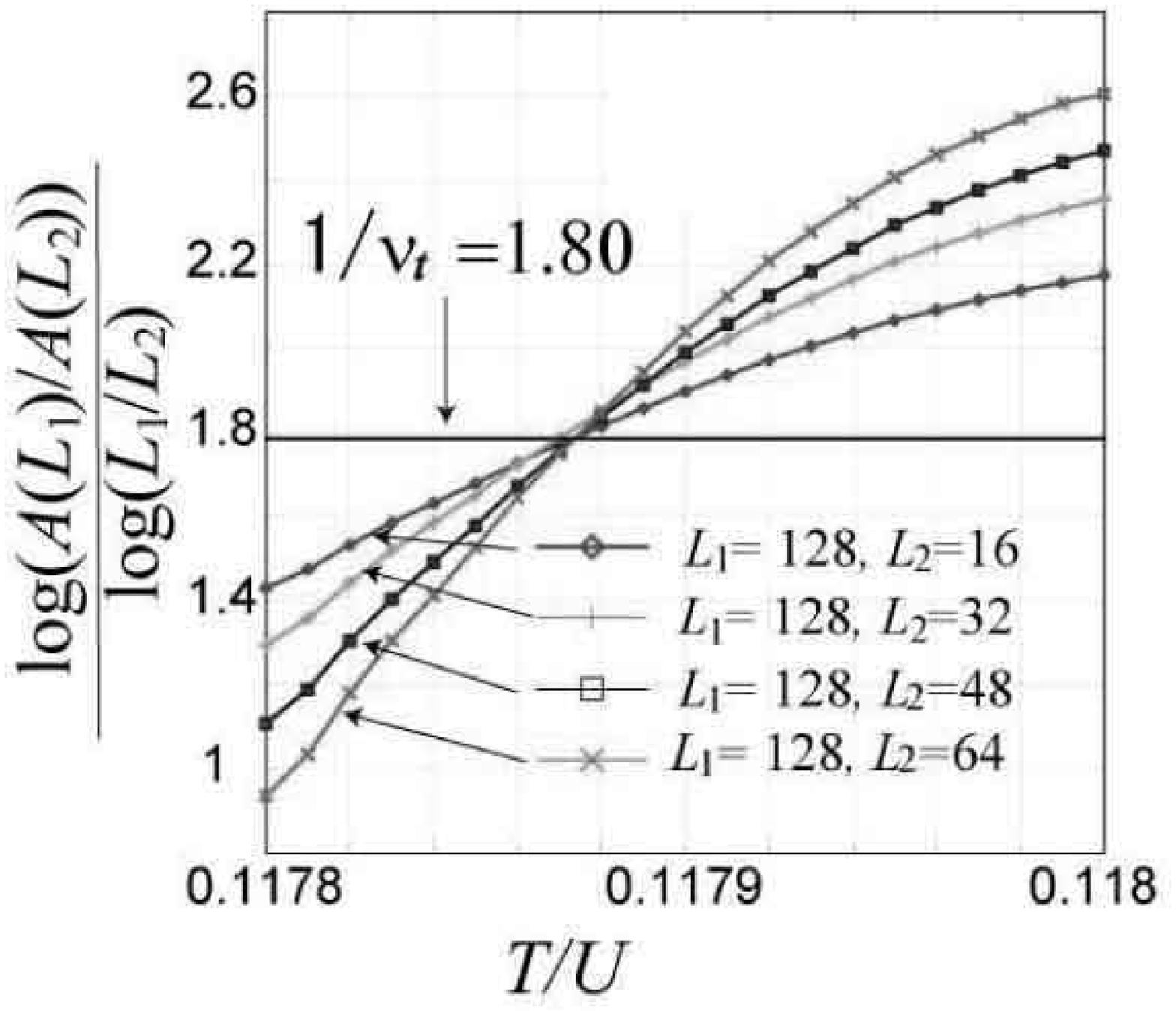}   
	\end{center}
\caption{Finite-size scaling of $A$ near TCP.
The parameters are  $L_{1}=128, L_{2}=16$ (open diamonds),
 $L_{1}=128, L_{2}=32$($+$),
 $L_{1}=128, L_{2}=48$ (squares),
$L_{1}=128, L_{2}=64$($\times$).
The horizontal line  shows the critical exponent in Ref. 10.}
\label{fig:TCP_nu}
\end{figure}

	We next consider the criticalities of TCP.
We perform finite-size scaling for $M$ and $A$. We determine the location of
TCP by the best fit of these finite-size scaling. We obtain the best fit for
$M$ and $A$ at $V/U=0.2545$. 
By the finite-size scaling of $M$ (Fig. \ref{fig:TCP_eta}), 
we estimate the critical temperature $T_{c}/U=0.111790(1)$ and the critical exponent 
$\eta_{t}=0.11(2)$. By the finite-size scaling of $A$ (Fig. \ref{fig:TCP_nu}),
we estimate $T_{c}/U=0.111788(1)$ and the critical exponent $\nu_{t}=0.55(3)$. 
These tricritical exponents are consistent with those obtained 
by the Monte Carlo renormalization group (MCRG) method~\cite{MCRG}.
Results by the MCRG method in the literature~\cite{MCRG} shows  $\eta_{t}=0.14(2)$, 
$\nu_{t}=0.56(1)$.

We summarize the critical exponents $\nu$ and $\eta$ obtained by Monte Carlo calculations
in Table \ref{CP}.
\begin{table}
\begin{center}
	\begin{tabular}{|l|c|c|} \hline
		\multicolumn{1}{|c|}{} &  \text{$\nu$} & \text{$\eta$} \\ \hline
		\multicolumn{1}{|c|}{\text{Lambda line}} & \text{$0.25(1)$} & \text{$1.03(5)$}  \\ \hline
 		\multicolumn{1}{|c|}{\text{TCP}} & \text{$0.55(3)$} & \text{$0.11(2)$}  \\ \hline
	\end{tabular}
\end{center}
\caption{Critical exponents $\nu$ and $\eta$ on the lambda line and at TCP.
The values in parentheses
denote the error bars in the last digit.}
\label{CP}
\end{table}%

\subsubsection{Various singularities and exponents}
	We now clarify the nature of fluctuations that diverge at TCP and the lambda line.
In this section, we will consider five fluctuations:  
$\chi_{m}$, $C$, $\chi_{D(\mathbf{0})}$, $\chi_{R}$
and $\chi_{c}$.

At TCP, the critical exponents $\gamma_{t}$ and $\gamma_{2t}$
are defined by
\begin{align}
	\chi_{m}&\sim |T-T_{c}|^{-\gamma_{t}} \label{eq:def_gamma_t}\\
	\chi_{D(\bf{0})}&\sim |T-T_{c}|^{-\gamma_{2t}}. \label{eq:def_gamma_2t}
\end{align}
Using the hyperscaling relation in Table \ref{HS_TCP}, 
we estimate  $\gamma_{t}$ and $\gamma_{2t}$ as
\begin{align}
	\gamma_{t}&=(2-\eta_{t})\nu_{t}=1.04(6) \label{eq:gamma_t}\\
	\gamma_{2t}&=2-2\nu_{t}=0.90(6). \label{eq:gamma_2t}
\end{align}
In Monte Carlo calculations,
we will show that other fluctuations 
$C$, $\chi_{R}$ and $\chi_{c}$ has the same singularity
 as that of $\chi_{D({\bf 0})}$ in general. 
We consider $\chi_{c}$ at half filling separately in the next section.

{\bf Lambda line: $\mu/U=1.2$}
	
 On the lambda line, the universality class of the phase transition is described by that of the 2D Ising model. 
Therefore, $\chi_{m}$ behaves as $\chi_{m} \sim |T-T_{c}|^{-\gamma}, \gamma=7/4$,
and  the singularity of the specific heat is logarithmic, $C \sim \log{|T-T_{c}|}$. 
This logarithmic singularity of  the specific heat  causes logarithmic singularities
of other  fluctuations  such as $\chi_{D(\mathbf{0})}$ and $\chi_{R}$  as proven by
 Ehrenfest law (see Appendix A).  From eqs. (\ref{eq:AP_C}), (\ref{eq:AP_D}), (\ref{eq:AP_R}) and (\ref{eq:AP_c}), 
$\chi_{D(\mathbf{0})}$, $\chi_{R}$
and $\chi_{c}$ diverge as $\log|T-T_{c}|$.

{\bf TCP: $\mu/U=1.2$}

From the hyperscaling relation in Table \ref{HS_TCP},
it is clear that the specific heat $C$
has the same critical exponent as that of the doublon 
density susceptibility $\chi_{D({\bf 0})}$. However,
the singularities of $\chi_{c}$ and $\chi_{R}$ 
are still not clear.

To apply Ehrenfest law, it is necessary
to  define partial derivatives. At the endpoint of
the continuous transition line, for example at TCP,
partial derivatives are not well defined. 
Therefore, at TCP, Ehrenfest law does not apply in general.
From this, it is nontrivial whether 
$\chi_{R}$ and $\chi_{c}$ have the same singularity 
as that of $\chi_{D({\bf 0})}$.

Near TCP, we confirm that
the doublon susceptibility $\chi_{D(\mathbf{0})}$, $\chi_{R}$,
the specific heat $C$, 
and the charge susceptibility $\chi_{c}$  are enhanced.
Using finite-size scaling for $\chi_{D(\mathbf{0})}$,
$\chi_{R}$, $C$ and $\chi_{c}$,
we estimate critical temperatures and the critical exponent $\gamma_{2t}/\nu_{t}$ as shown in 
Table \ref{TCP}. 
These results are consistent with the value estimated 
from the hyperscaling relation in Table \ref{HS_TCP} as
\begin{equation}
	\gamma_{2t}/\nu_{t}=2/\nu_{t}-2=1.6\pm 0.1.
\end{equation}
From these results, we conclude that $\chi_{R}$ and $\chi_{c}$ have the same singularity
as that of $\chi_{D({\bf 0})}$.
The positive $\gamma_{2t}$ indicates that $\chi_{D}$, $\chi_{R}$, 
$C$ and $\chi_{c}$ diverge at TCP.

\begin{table}
\begin{center}
	\begin{tabular}{|l|c|c|} \hline
		
		\multicolumn{1}{|c|}{} & \text{$T_{c}$} & \text{exponent} \\ \hline

		\multicolumn{1}{|c|}{\text{$\chi_{D(\mathbf{0})}$} }
		& \text{$0.11786(3)$} &  \text{$1.7(2)$}\\ \hline 
		
		\multicolumn{1}{|c|}{\text{$\chi_{R}$}}
	   	&  \text{$0.11786(3)$} & \text{$1.7(2)$} \\ \hline
		
       \multicolumn{1}{|c|}{\text{$C$}}
		 & \text{$0.11785(3)$} & \text{$1.5(2)$}  \\ \hline
		 
	\multicolumn{1}{|c|}{\text{$\chi_c$}} 
	&  \text{$0.11784(2)$} & \text{$1.4(2)$} \\ \hline
		
		\multicolumn{1}{|c|}{\text{$\frac{\gamma_{2t}}{\nu_{t}}$}}
		&   &\text{$1.6(1)$} \\ \hline
	\end{tabular}
\end{center}
\caption{Critical temperatures and critical exponents at TCP
obtained from Monte Carlo calculations. The values in parentheses 
denote the error bars in the last digit.}
\label{TCP}
\end{table}%

\subsubsection{Charge susceptibility: half filling}
\label{chi_c}
{\bf Lambda line}

\begin{figure}[htp]
	\begin{center}
		\includegraphics[width=6cm]{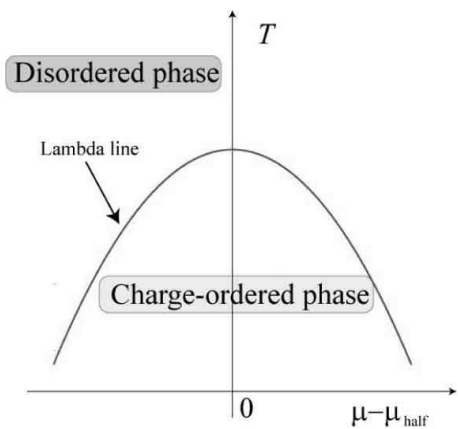}   
	\end{center}
\caption{Phase diagram in the $\mu$-$T$ plane for $V/U>0.2546$. 
$\mu_{half}=4V+U/2$ is the chemical 
potential that corresponds to the case of half filling. The transition point at half filling
is at the peak of the lambda line in the $\mu$-$T$ plane.}
\label{fig:PH_sym_1}
\end{figure}

At half filling, $\chi_{c}$ is not enhanced near the lambda line, while other fluctuations are enhanced.
This is due to the particle-hole symmetry. Because of the particle-hole symmetry,
the phase diagram should be
symmetric in the $\mu$-$T$ plane, as shown in Fig. \ref{fig:PH_sym_1}.
Half filling corresponds to the peak of the lambda line and the slope
of the lambda line at half filling is zero. 
From the condition that the slope of the lambda line in the $\mu$-$T$ plane is zero
and  eqs. (\ref{Ehren_1}) and (\ref{Ehren_2}),
$\chi_{c}$ has a singularity weaker than that of the specific heat $C$
leading to a singularity weaker than the logarithmic one on the lambda line,
and we do not observe the enhancement of $\chi_{c}$ on the lambda line.

{\bf TCP}

At half filling, we estimate the location of TCP at $T_{c}/U=0.1567$
and $V_{c}/U=0.2546$ by the best fit for $M$ and $A$ in eqs. (\ref{eq:eta}) 
and (\ref{eq:nu}).

\begin{figure}[htp]
	\begin{center}
		\includegraphics[width=6cm]{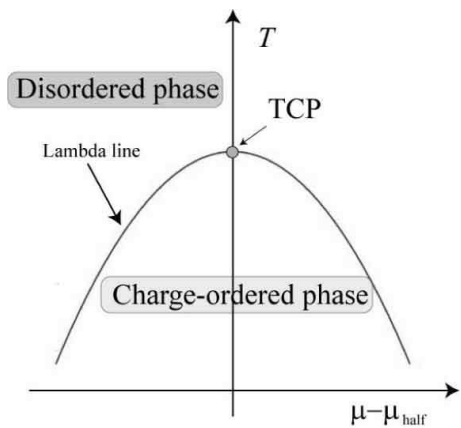}   
	\end{center}
\caption{Phase diagram in the $\mu$-$T$ plane at $V/U=0.2546$,
which contains TCP.
TCP at half filling is located at
the peak of the lambda line in the $\mu$-$T$ plane.}
\label{fig:PH_sym_2}
\end{figure}

\begin{figure}[htp]
	\begin{center}
		\includegraphics[width=6cm]{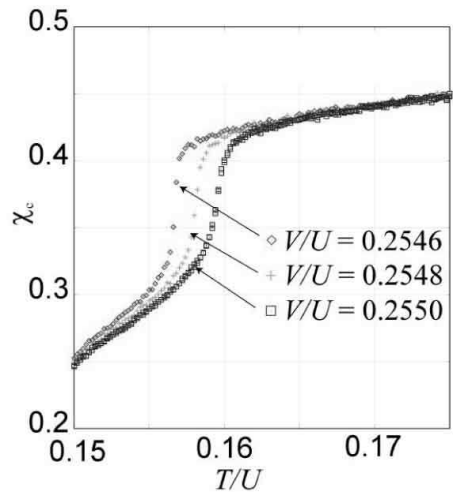}   
	\end{center}
\caption{$\chi_c$ near TCP ($V_{c}/U=0.2546$, $T_{c}/U=0.1567$) at half filling.
The parameters are  $V/U=0.2546$ (open diamonds),
$V/U=0.2548$($+$), $V/U=0.2550$ (squares) and the linear dimension of 
the lattice size is $L=128$. We do not find the divergence of $\chi_{c}$ up to $L=128$. }
\label{fig:CHI_C_HALF_MC}
\end{figure}

As mentioned above, Ehrenfest law does not apply to TCP in general.
However, only at half filling  Ehrenfest law can be applied 
to TCP, because TCP is not the endpoint of the lambda line 
in the $\mu$-$T$ plane as shown in Fig. \ref{fig:PH_sym_2}.
Using the same reasoning for the lambda line, $\chi_{c}$ has a singularity
weaker than that of the specific heat $C$. Therefore $\chi_{c}$ does not diverge with the
singularity $|T-T_{c}|^{-\gamma_{2t}}$ and we do not find  
the divergence of $\chi_{c}$ near TCP at half filling, as shown in Fig. \ref{fig:CHI_C_HALF_MC}.  

We have clarified which fluctuations diverge at TCP in both 
mean-field approximation and Monte Carlo calculation.
We confirm that such divergences occur in the same quantities
between the mean-field approximation and the Monte Carlo results.
Exact calculations give only modifications of the critical exponents
from those of the mean-field results.
Except for the charge susceptibility $\chi_{c}$,
all the fluctuations we consider diverge with the power law at TCP.
Because of the particle-hole symmetry, 
$\chi_{c}$ does not happen to diverge at half filling.
However, away from half filling, we confirm that $\chi_{c}$ generically diverges at TCP
with the same singularity as that of  $\chi_{D(\bf{0})}$. 
Finally, the critical exponents we obtain in the classical model is shown in Table \ref{Crex}. 
We note that various susceptibilities and the specific heat diverge much more strongly
than those at the lambda transition. 
We have for the first time shown that the critical exponent of the charge susceptibility
$\chi_{c}$ and the susceptibility of the next-nearest correlation $\chi_{R}$ is the 
same as that of the susceptibility of $\chi_{D({\bf 0})}$ in general. 
It is remarkable  that some fluctuations at zero
wavenumber diverge as $\chi_{D(\mathbf{0})}$ 
in contrast to the lambda line.
\begin{table}
\begin{center}
	\begin{tabular}{|l|c|c|c|c|} \hline
		\multicolumn{1}{|c|}{} & \text{MF $\lambda$-line} 
		& \text{MF TCP}
		 & \text{MC $\lambda$-line}
		 & \text{MC TCP} 	\\ \hline
		\multicolumn{1}{|c|}{\text{$m$}}
		 & \text{$\beta=\frac{1}{2}$} 
		& \text{$\beta_{t}=\frac{1}{4}$}
		 & \text{$\beta=0.12(6)$}
		 & \text{$\beta_{t}=0.030(7)$} 	\\ \hline
		\multicolumn{1}{|c|}{\text{$\langle D({\bf 0}) \rangle$}}
		 & \text{$\beta_{2}=1$} 
		& \text{$\beta_{2t}=\frac{1}{2}$}
		 & \text{$$}
		 & \text{$\beta_{2t}=0.10(6)$} 	\\ \hline
		\multicolumn{1}{|c|}{\text{$\chi_{m}$}}
		 & \text{$\gamma=1$} 
		& \text{$\gamma_{t}=1$}
		 & \text{$\gamma=1.8(1)$}
		 & \text{$\gamma_{t}=1.04(4)$} 	\\ \hline
		\multicolumn{1}{|c|}{\text{$\chi_{D({\bf 0})}$}}
		 & \text{$\gamma_{2}=0$} 
		& \text{$\gamma_{2t}=\frac{1}{2}$}
		 & \text{$\log$}
		 & \text{$\gamma_{2t}=0.9(2)$} 	\\ \hline
		\multicolumn{1}{|c|}{\text{$C$}}
		 & \text{$\alpha=0$} 
		& \text{$\alpha_{t}=\frac{1}{2}$}
		 & \text{$\log$}
		 & \text{$\alpha_{t}=0.8(2)$} 	\\ \hline
		\multicolumn{1}{|c|}{\text{$\chi_{R}$}}
		 & \text{$0$} 
		& \text{$\frac{1}{2}$}
		 & \text{$\log$}
		 & \text{$\text{0.9(2)}$} 	\\ \hline
		\multicolumn{1}{|c|}{\text{$\chi_{c}$}}
		 & \text{cusp} 
		& \text{cusp*}
		 & \text{$\log $**}
		 & \text{$\text{0.8(2)}$**} 	\\ \hline
	\end{tabular}
\end{center}
\caption{Critical exponents at lambda line and TCP. MF and MC represent
mean-field results and Monte Carlo results, respectively. 
The mean-field results are obtained at half filling and
the Monte Carlo results are obtained in the case at $\mu/U=1.2$.
The values in parentheses denote the error bars in the last digit.
$^{*}$ Away from half filling ($\mu/U=1.2$), $\chi_{c}$ diverges
with the singularity $\chi_{c}\sim |V-V_{c}|^{-0.499}$. 
$^{**}$At half filling,
$\chi_{c}$ does not diverge on the lambda line and at TCP.}
\label{Crex}
\end{table}%

\section{Itinerant models}
\label{Itinerant}
\subsection{Model}
In this section, we consider the effect of itinerancy of particles on
TCP within the Hartree-Fock approximation. We add the hopping term to the classical model.
In this section, we study charge ordering transition as well as the competition
between metals and insulators. Metal-insulator transitions are also discussed in terms of the singularity 
of conductivity. We first consider the extended Hubbard model with the next-nearest-neighbor hopping($t^{\prime}$) on
the two-dimensional square lattice. The Hamiltonian is given by eq. (\ref{EHM}).

\subsection{Hartree-Fock approximation}
We consider checker-board-type doublon alignment described by the order parameter $m$
defined from $m=1/N_{s}\sum_{i}\langle (n_{i\uparrow}+n_{i\downarrow})\exp(i\mathbf{Q}\mathbf{r_{i}})\rangle$.
Here, we take $\langle n_{i\uparrow}\rangle=\langle n_{i\downarrow}\rangle=(n+m\exp(i\mathbf{Q}\mathbf{r_{i}}))/2$,
with $n$ being the average  charge density given by $1/N_{s}\sum_{i}\langle n_{i\uparrow}+n_{i\downarrow}\rangle$.
Within the Hartree-Fock approximation, we decouple the interaction term.

The onsite interaction term is decoupled as 
\begin{align}
	&U\sum_{i}n_{i\uparrow}n_{i\downarrow}  \\  \notag
&=U\sum_{i}\left(n_{i\uparrow}-\langle n_{i\uparrow}\rangle+\langle n_{i\uparrow}\rangle\right)
\left(n_{i\downarrow}-\langle n_{i\downarrow}\rangle+\langle n_{i\downarrow}\rangle\right) \\  \notag
&\sim U\sum_{i}\left(n_{i\uparrow}\langle n_{i\downarrow}\rangle+n_{i\downarrow}\langle n_{i\uparrow}\rangle
-\langle n_{i\uparrow}\rangle\langle n_{i\downarrow}\rangle\right) \notag \\ 
&=\frac{U}{2}\sum_{i\sigma}\left(n_{i\sigma}\left(n+me^{i\mathbf{Qr_{i}}}\right)\right)-\frac{UN_{s}\left(n^2+m^2\right)}{4}. \notag \\
\end{align}

The nearest-neighbor interaction term is decoupled as 
\begin{align}	
	&V\sum_{\langle ij\rangle}N_{i}N_{j} \\  \notag
	&=V\sum_{\langle ij\rangle}\left((N_{i}-\langle N_{i}\rangle
	+\langle N_{i}\rangle)(N_{j}-\langle N_{j}\rangle+\langle N_{j}\rangle)\right) \notag \\
    &\sim V\sum_{\langle ij\rangle}(N_{i}\langle N_{j}\rangle
	+N_{j}\langle N_{i}\rangle-\langle N_{i}\rangle\langle N_{j}\rangle) \notag \\
  &=4nV\sum_{i}N_{i}-4mV\sum_{ i}N_{i}\exp(i\mathbf{Qr_{i}})-2VN_{s}(n^2-m^2). 
\end{align}

Finally, we obtain a Hartree-Fock Hamiltonian ($H_{HF}$) in momentum space as 
\begin{align}
 H_{HF}&=\sum_{k,\sigma}\left(\epsilon(k)+\frac{Un}{2}+4nV\right)c_{\mathbf{k} \sigma}^{\dagger}c_{\mathbf{k} \sigma} \\  \notag
	&+\left(\frac{U}{2}-4V\right)m\sum_{\mathbf{k} \sigma}c_{\mathbf{k+Q} \sigma}^{\dagger}c_{\mathbf{k} \sigma} \notag \\	
 	&-\left(\frac{UN_{s}(n^2+m^2)}{4}+2VN_{s}(n^2-m^2)\right),
\end{align}
where $\epsilon(k)=-2t(\cos(k_x)+\cos(k_y))+4t^{\prime}\cos(k_x)\cos(k_y)$.

Diagonalizing the Hamiltonian leads to two bands of the form
\begin{align}
E_{\pm}&=\frac{\epsilon(\mathbf{k})+\epsilon(\mathbf{k+Q})}{2}+\frac{Un}{2}+4nV \\  \notag
&\pm \sqrt{(\frac{\epsilon(\mathbf{k})-\epsilon(\mathbf{k+Q})}{2})^{2}+(mg)^{2}}, \label{eq:Band}
\end{align}
where $g=4V-\frac{U}{2}$. 

Using this Hartree-Fock band dispersion, we obtain the free energy 
\begin{align}
 F_{HF}&=-\frac{T}{N_{s}}\log Z_{HF} \notag \\  
	&=-\frac{2T}{ N_{s}}\sum_{\mathbf{k},\eta}\log(1+e^{-(E_{\eta}(\mathbf{k})-\mu)/T})] \notag \\
	&-\left(\frac{U(n^2+m^2)}{4}+2V(n^2-m^2)\right),
\end{align}
where the suffix $\eta$ takes $\pm$.

	From this free energy, the charge density $n$ and the order parameter $m$ are determined from the
self-consistency condition 
\begin{align}
	n&=\frac{2}{N_{s}}\sum_{\mathbf{k}}f\left(E_{+}(\mathbf{k})\right)+f\left(E_{-}(\mathbf{k})\right)  \\
	\frac{1}{g}&=2\sum_{\mathbf{k}} \frac{f(E_{-}(\mathbf{k}))-f(E_{+}(\mathbf{k}))}{E_{+}(\mathbf{k})-E_{-}(\mathbf{k})}, \label{eq:HF_self}
\end{align}
where $f(x)$ is the Fermi-Dirac distribution function.

	We also obtain the doublon density $\langle D(\mathbf{0})\rangle$ and $R$ conjugate to $V$ defined by
\begin{align}
	\langle D(\mathbf{0})\rangle&=\langle n_{i\uparrow}\rangle\langle n_{i\downarrow}\rangle=\frac{n^2+m^2}{4}, \notag \\
	R&=\langle N_{i}\rangle\langle N_{j}\rangle=2(n^2-m^2). 	
\end{align}
 	$\langle D(\bf{0})\rangle$ and $R$ are described by the 
square of $m$, which are notable features of the  mean-field approximation.

It turns out that the self-consistent equation (\ref{eq:HF_self}) with 
eq. (\ref{eq:Band}) for the extended Hubbard model (\ref{EHM})  is equivalent
to the self-consistent equation for the simple Hubbard model obtained by taking
$V=0$ in eq. (\ref{EHM}).
The equivalence holds by replacing the charge-order parameter
$m$ with the antiferromagnetic(AF) order $m_{AF}=\sum_{i}\langle (n_{i\uparrow}-n_{i\downarrow})\exp({\bf Qr_{i}})\rangle/N_{s}$
and by putting  $g=U/2$. 
However, one should be careful about this mapping.
This mapping is complete only at the Hartree-Fock level.
The charge order is the consequence of discrete symmetry breaking 
while the AF order is realized by the continuous symmetry breaking of $SU(2)$ 
symmetry. Therefore, if it would be exactly solved  
in two-dimensional systems, the charge order can indeed exist at finite temperatures, though 
AF order cannot exist at finite temperatures by Mermin-Wargner theorem.
Therefore, the Hartree-Fock approximation captures 
some essence of the  charge order at finite temperatures,
while the AF order at nonzero temperatures in two dimensions is 
an artifact of the Hartree-Fock approximation.	

\subsection{Critical exponents}
 We first consider the case with the electron density fixed at $n=1$ in the canonical ensemble.
We determine the location of TCP, and the critical exponent of the order parameter.
The behavior of the order parameter $m$ at TCP is different from  that of the lambda line as 
\begin{equation}
	m\propto |g-g_{c}|^{\beta} : \begin{cases}
									&\beta=\frac{1}{2} \text{ for lambda line} \notag \\												
									&\beta_{t}=\frac{1}{4} \text{ for TCP}	.
								\end{cases}	
\end{equation}

\begin{figure}[htp]
	\begin{center}
		\includegraphics[width=5cm]{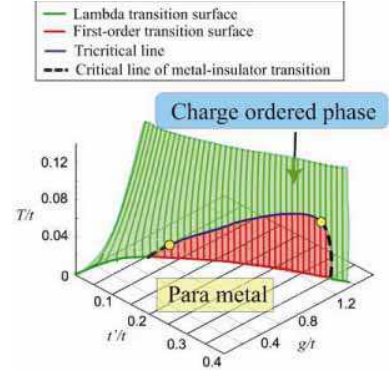}   
	\end{center}
\caption{Phase diagram in $T$-$g$-$t^{\prime}$  space.
The lambda (first-order) surface
is described as green (red) surface above (below) 
solid and broken lines, respectively.
The TCP line is shown in blue. Near zero temperature,
at the yellow circle, the TCP line goes into the ordered phase and changes into 
the critical line of the metal-insulator transition
(broken black line).} 
\label{fig:G_Phase}
\end{figure}

\begin{figure}[htp]
	\begin{center}
		\includegraphics[width=5cm]{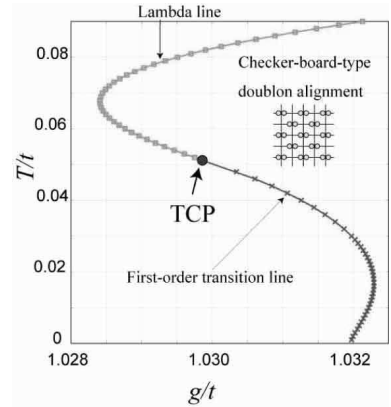}   
	\end{center}
\caption{Phase diagram at $t^{\prime}/t=0.2$.
We estimate the location of TCP as
 $T_{c}/t=0.0511$ and  $g_{c}/t=0.01030$}.
\label{fig:HF}
\end{figure}

\begin{figure}[htp]
	\begin{center}
		\includegraphics[width=5cm]{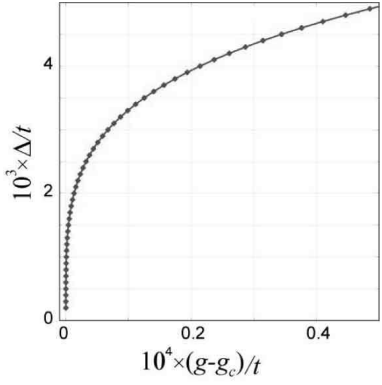}   
	\end{center}
\caption{Singularity of the order parameter at TCP.
$\Delta$ is defined by $\Delta=mg$, which has the same singularity  
as that of the order parameter $m$. We estimate tricritical temperature as $T_{c}/t=0.0511$
and tricritical interaction as $g_{c}/t=0.01030$}
\label{fig:HF_TCP_order}
\end{figure}

\begin{figure}[htp]
	\begin{center}
		\includegraphics[width=6cm]{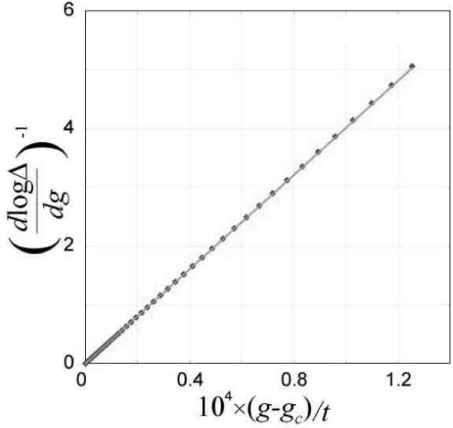}   
	\end{center}
\caption{Critical exponent of the order parameter at TCP.
$\Delta$ has the singularity such that $\Delta=a|g-g_{c}|^{\beta_{t}}$.
Therefore we obtain $(d \log\Delta/d g)^{-1}=\frac{1}{\beta_{t}}|g-g_{c}|$.
Using this relation we estimate $\beta_{t}=0.249\pm0.002$ as the fitting shown
by the solid line.}
\label{fig:HF_TCP_scale}
\end{figure}

A schematic phase diagram is shown in Fig. \ref{fig:G_Phase}.
The TCP line appears at finite temperatures in $T$-$g$-$t^{\prime}$  space.
In this paper, we study TCP at $t^{\prime}/t=0.2$ in detail. 
The phase diagram at $t^{\prime}/t=0.2$ is shown in Fig. \ref{fig:HF}.
The singularity of the order parameter is shown in Fig. \ref{fig:HF_TCP_order} and 
scaling result is shown in Fig. \ref{fig:HF_TCP_scale}. The critical exponent of the order parameter
 $\beta_{t}=0.249\pm0.002$ is well consistent with the expected mean-field value $\beta_{t}=1/4$. 

 We specify which fluctuations diverge at TCP and on the lambda line 
in the itinerant model (\ref{EHM}).
The singularity of the doublon susceptibility $(\chi_{D(\mathbf{0})})$ 
and fluctuation conjugate to the nearest-neighbor repulsion 
V $(\chi_{R})$ are respectively given as 
\begin{align}
	\chi_{D(\mathbf{0})}=\frac{\partial \langle D(\mathbf{0})\rangle}{\partial U}\sim |g-g_{c}|^{2\beta-1}		\notag \\
	\chi_{R}=\frac{\partial R}{\partial U}\sim |g-g_{c}|^{2\beta-1}.		\notag
\end{align}
At TCP, since $\beta_{t}=1/4$, $\chi_{D(\bf{0})}$ and
$\chi_{R}$ respectively behave as

\begin{align}
	\chi_{D(\mathbf{0})}\sim |g-g_{c}|^{-\frac{1}{2}}		\notag \\
	\chi_{R}\sim |g-g_{c}|^{-\frac{1}{2}}.		\notag
\end{align}
This means that $\chi_{D(\mathbf{0})}$ and $\chi_{R}$ diverge at TCP,
in contrast to the criticality on the lambda line. 

The singularity of the charge susceptibility will be discussed in \S \ref{sec:PS}. 

\subsection{Metal-insulator transition}
\begin{figure}[htp]
	\begin{center}
		\includegraphics[width=5cm]{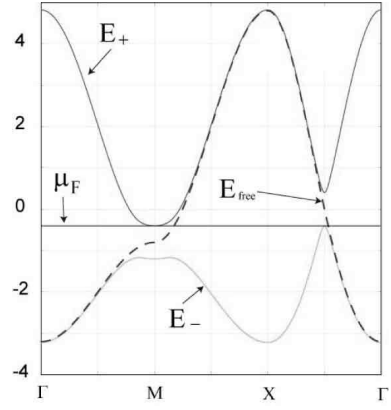}   
	\end{center}
\caption{Band dispersions at the transition point at zero temperature are shown.
The Fermi energy for the charge-ordered side is described by $\mu_{F}$. Band dispersions are given by 
$E_{\pm}=4t^{\prime}\cos(k_{x})\cos(k_{y})\pm\sqrt{4t^{2}(\cos(k_{x})+\cos(k_{y}))^{2}+\Delta_{c}^{2}}$
for the charge-ordered phase
and $E_{\rm{free}}=-2t(\cos(k_{x})+\cos(k_{y}))+4t^{\prime}\cos (k_{x})\cos(k_{y})$ 
for the parametal phase
where $\Delta_{c}/t=0.4031$.}
\label{fig:Band}
\end{figure}

We consider the relation between metal-insulator transitions  
and charge-order transitions at $t^{\prime}/t=0.2$ in this section.
At zero temperature, the metal-insulator transition and
the charge-order transition occur at the same time. 
Namely, the charge gap opens for the whole Brillouin-zone 
simultaneously with the charge-order transition. The critical $\Delta$ of the
metal-insulator transition at $t^{\prime}/t=0.2$ is given by
$\Delta_{c}/t=2t^{\prime}/t=0.4$.
The gap $\Delta$ shows a jump from $\Delta/t=0$ to $\Delta/t=0.4031$
at the first-order transition point at $g/t=1.0319$.
Figure \ref{fig:Band} shows the band dispersions of the parametal side and 
charge-ordered side at the transition point.

\begin{figure}[htp]
	\begin{center}
		\includegraphics[width=5cm]{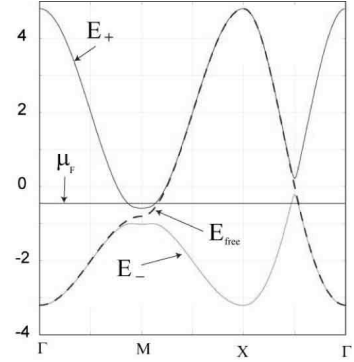}   
	\end{center}
\caption{Band dispersions at the transition point at $T/t=0.04$ are shown.
The Fermi energy for the charge-ordered side is described by $\mu_{F}$.
Band dispersions are given by 
$E_{\pm}=4t^{\prime}\cos (k_{x})\cos(k_{y})\pm\sqrt{4t^{2}(\cos(k_{x})+\cos(k_{y}))^{2}+\Delta_{c}^{2}}$
and $E_{\rm{free}}=-2t(\cos(k_{x})+\cos(k_{y}))+4t^{\prime}\cos (k_{x})\cos(k_{y})$
 where $\Delta_{c}/t=0.2170$.}
\label{fig:Band_T40}
\end{figure}

At finite temperatures, the jump of $\Delta$ decreases and vanishes at TCP ($T_{c}/t=0.0511$).
Therefore, the metal-insulator transition, strictly speaking metal-semimetal transition, 
and the charge-order transition occur separately at high temperatures.
Figure \ref{fig:Band_T40} shows the band dispersions in the parametal side,
$E_{\rm{free}}$ and  in the charge-ordered side, $E_{\pm}$
at the transition point at $T/t=0.04$. The gap $\Delta$ shows a jump from $\Delta/t=0$ to $\Delta/t=0.2170$
at the first-order transition point at $g/t=1.0310$. Therefore, the charge gap does not fully 
open for the whole Brillouin zone.

\subsection{Conductivity}
 In this section, we consider the singularity of carrier density.
The definition of carrier density is given by the electron density in the upper band as
\begin{equation}
	X=\int_{BZ}f(E_{+}(\mathbf{K}))d\mathbf{k}. \label{eq:Carrier_Density}
\end{equation} 
Phenomenologically the conductivity may be expressed by the product
of the carrier density $X$ and the carrier relaxation time $\tau$ 
as $\sigma\propto X\tau$. In this section, we consider the
singularity of the carrier density $X$. The singularity 
of the relaxation time will be considered later. 

First, we consider how the carrier density depends on the gap $\Delta$. In our calculation,
we first fix  $\Delta$ and determine the interaction $g$ using the self-consistent 
equation. This allows us to determine the relation between the carrier density and $\Delta$ 
even in the first-order transition region. 
From eq. (\ref{eq:Log_correct}) in Appendix B,
the singularity of carrier density is given as
$|X-X_{c}|=\Delta^{2}(A\log{\Delta}+B)$ in the asymptotic region.
This relation holds both near the lambda transition as well as near TCP. 

On the lambda line, the conductivity exponent $p$ defined by 
$|X-X_{c}|\propto (g-g_{c})^{p}(A_{\Delta}\log{\Delta}+B_{\Delta})$ is given
from $p=2\beta=1$ since $\Delta\propto (g-g_{c})^{\beta}$ with $\beta=1/2$.

At TCP in the canonical ensemble, the conductivity exponent $p_{t}$
defined by $|X-X_{c}|\propto (g-g_{c})^{p_{t}}(A\log{\Delta}+B)$ is given
from $p_{t}=2\beta_{t}=1/2$ since $\Delta\propto (g-g_{c})^{\beta_{t}}$ with $\beta_{t}=1/4$.
As we will show later, at TCP in the grand-canonical ensemble,
the critical exponent $\beta_{t}=1/4$ changes into $\beta=1/2$.
Therefore the conductivity exponent $p_{t}$ changes into $p_{t}^{\prime}=2\beta=1$.
The exponent $p$ on the lambda line is twice as large as $p_{t}$ 
at TCP in the canonical ensemble.

\begin{figure}[htp]
	\begin{center}
		\includegraphics[width=6cm]{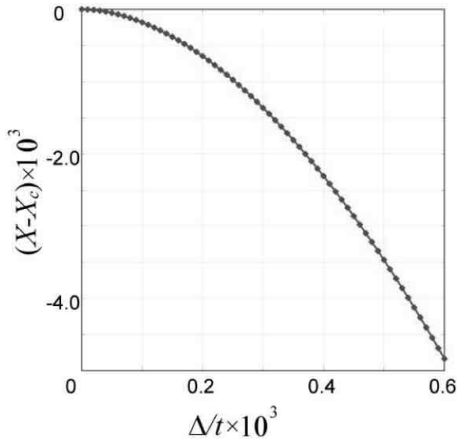}   
	\end{center}
\caption{Carrier density $X$ is plotted near TCP.}
\label{fig:CP_Con}
\end{figure}

\begin{figure}[htp]
	\begin{center}
		\includegraphics[width=5cm]{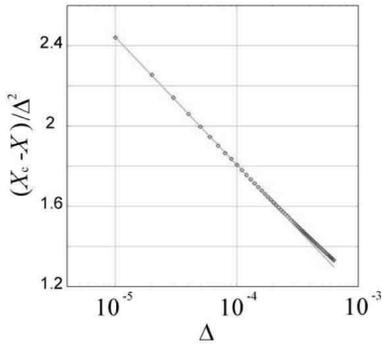}   
	\end{center}
\caption{Singularity of $(X_{c}-X)/\Delta^{2}$ is plotted near TCP
($T/t$=0.0511) as (red) diamonds.  We confirm that $(X_{c}-X)/\Delta^{2}$ has the singularity
$A\log{\Delta}+B$ shown by the solid line, where 
we estimate $A=-0.276(2)$ and $B=-0.73(2)$.}
\label{fig:TCP_CARRIER-GAP_1}
\end{figure}

Numerically, near TCP, we confirm that carrier density has the singularity defined 
in eq. (\ref{eq:Log_correct}). The singularity of the carrier density is shown 
in Fig. \ref{fig:CP_Con}.
We fit $(X_{c}-X)/\Delta^{2}$ with 
the function $A\log{\Delta}+B$, then obtain $A=-0.276(2)$ and $B=-0.73(2)$.
This result is shown in Fig. \ref{fig:TCP_CARRIER-GAP_1}.

\subsection{Phase Separation}
\label{sec:PS}
\begin{figure}[htp]
	\begin{center}
		\includegraphics[width=10cm]{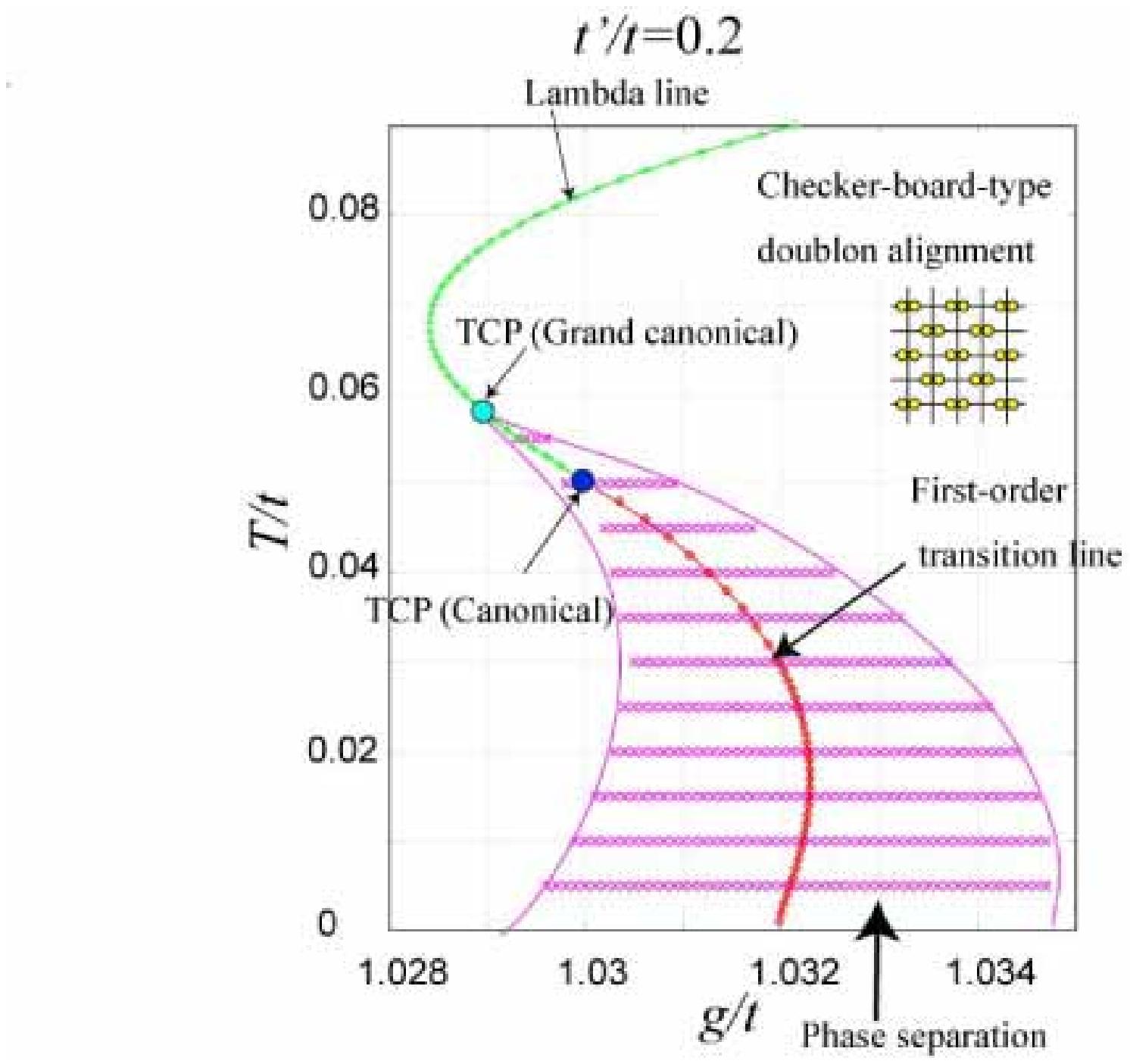}   
	\end{center}
\caption{Phase separation region in the case of
$U/t=10$ is described as pink hatched region.
TCP in the canonical ensemble is unstable and TCP in the grand-canonical
ensemble shifts to the point at $T_{c}/t=0.05726$, $g_{c}/t=1.029046$.}
\label{fig:P_Sep_1}
\end{figure}

In this section, we consider the phase separation near TCP in the grand-canonical
ensemble.
In our model, we confirm that the phase separation occurs at $U/t=10.0$ not only along the
first-order transition line but also along the second-order transition line
as shown in Fig. \ref{fig:P_Sep_1}.
Therefore TCP in the canonical ensemble is actually 
unstable toward the phase separation if the grand-canonical ensemble is employed.
If we calculate in the grand-canonical ensemble, the location of TCP shifts from  
the canonical one. 

Here, we note that the region of the phase separation depends
on the onsite interaction $U$. 
The definition of $\mu$ is given as
\begin{align}
	\mu&=\mu_{F}+(4V+\frac{U}{2})n \notag \\ 
		&=\mu_{F}+(g+U)n
\end{align}
where $\mu_{F}$ is the Fermi energy and it only depends on temperature and  band structure, namely
$t$, $t^{\prime}$ and $\Delta$. Since $U$ is independent of $g$, $\mu$ depends on $U$ and the region of
the phase separation depends on $U$.
Roughly speaking, the phase separation region shrinks for larger
$U$ and above the possible threshold $U_{c}$, TCP of the
canonical ensemble becomes stable. 

\begin{figure}[htp]
	\begin{center}
		\includegraphics[width=6cm]{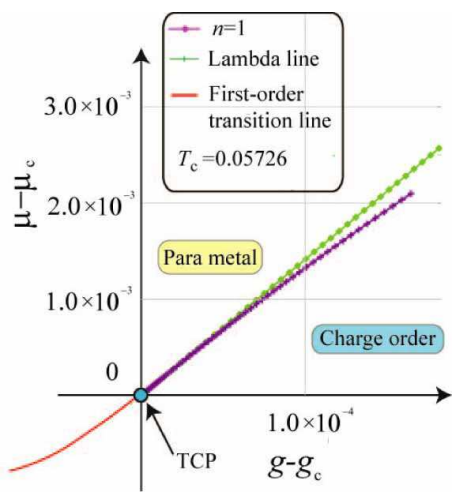}   
	\end{center}
\caption{The lambda line and the line which corresponds to the half filling
in the $\mu$-$g$ plane at the critical temperature $T_{c}/t=0.05726$ are plotted.
Near TCP, half-filling line becomes tangential to the lambda line. Therefore, the
singularity of the order parameter is given by $m\sim |g-g_{c}|^{\beta}$ where $\beta=1/2$.}
\label{fig:mu_g_TCP}
\end{figure}

At $U/t=10$, the location of TCP is at $T_{c}/t=0.05726$ and $g_{c}/t=1.02905$.
Because the half-filling plane is tangential to the lambda surface
in $T$-$\mu$-$g$ space near TCP as we see in Fig. \ref{fig:mu_g_TCP},
 the critical exponent of the order parameter
is different from the generic one of TCP but the same
as that of the lambda line. In the route that approaches
TCP along the lambda line, the critical exponents of the order parameter
and the charge susceptibility are respectively given by
\begin{align}
	m&\sim|g-g_{c}|^{\beta} \\
	\chi_{c}&\sim|g-g_{c}|^{-\gamma_{2}} 
\end{align}
where $\beta=1/2$ and $\gamma_{2}=1$~\cite{TCP}.
However, for the generic filling-control transition
toward TCP, it is away from the lambda line and the exponents
should recover the generic tricritical exponents $\beta_{t}=1/4$
and $\gamma_{2t}=1/2$.

Figure \ref{fig:mu_g_TCP} is the phase diagram
at the critical temperature $T_{c}/t=0.05726$. We obtain the singularity of 
the order parameter as $m\sim |g-g_{c}|^{0.499\pm0.002}$.
This value is consistent with the  mean-field value.
However, the exponent $p_{\Delta}$ is the same as that of the generic one.
Therefore, we obtain the singularity of the carrier density 
as $|X-X_{c}|\sim |g-g_{c}|^{p_{t}^{\prime}}$ where $p_{t}^{\prime}=2\times \beta=0.92\pm 0.01$.
In the present model, TCP appears to move out of the phase separation region
for $U/t\gtrsim 100$, and canonical and grand-canonical results give the same TCP.

In real materials, long-range Coulomb interaction plays an important role
and it can suppress the phase separation. Therefore, the TCP
in real materials may be compromised and located between the grand-canonical
and canonical TCP's. Critical exponents in this case are not
clear at the moment. Effects of the long-range
Coulomb interaction on the phase separation is left for future studies.

\begin{figure}[htp]
	\begin{center}
		\includegraphics[width=6cm]{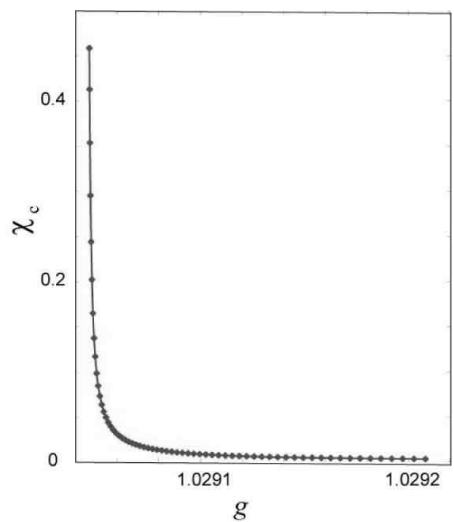}   
	\end{center}
\caption{$\chi_{c}$ near TCP in the grand-canonical ensemble is plotted.}
\label{fig:Chi_c_Half_HF}
\end{figure}

\begin{figure}[htp]
	\begin{center}
		\includegraphics[width=6cm]{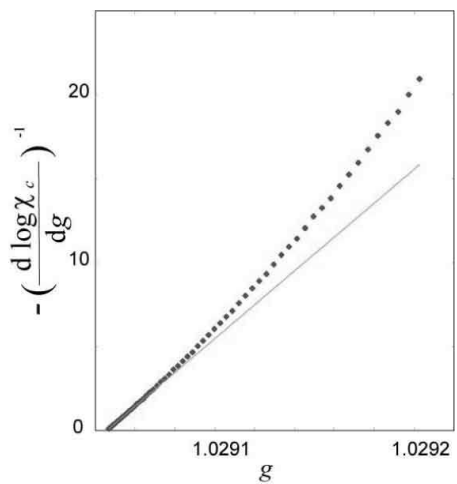}   
	\end{center}
\caption{Scaling analysis of $\chi_{c}$. Slope of green line
corresponds to the critical exponent  $\gamma_{2}$. We
estimate $\gamma_{2}=0.99\pm0.02$. This value is consistent with 
the mean-field value $\gamma_{2}=1$.}
\label{fig:Chi_c_Half_HF_SC}
\end{figure}

We calculate the charge susceptibility $\chi_{c}$ near TCP in the grand-canonical 
ensemble. Figure \ref{fig:Chi_c_Half_HF} shows the charge susceptibility $\chi_{c}$ near TCP at $T_{c}/t=0.05726$.
We obtain the singularity of charge susceptibility as $\chi_{c}\sim |g-g_{c}|^{-\gamma_{2}}$
where $\gamma_{2}=0.99\pm0.01$ as shown in Fig. \ref{fig:Chi_c_Half_HF_SC}. 
This value is consistent with the mean-field value $\gamma_{2}=1$~\cite{TCP}.

\section{Comparison with experimental results}
\label{Conductivity}
 In this section, we compare the present Hartree-Fock analysis with experimental results.
An organic conductor $\rm{(DI-DCNQI)_{2}Ag}$ is a compound
with the uniform stacking of planar DI-DCNQI molecules along
the $c$ axis forming conductive chains with monovalent and nonmagnetic Ag ions
stacked among the chains.
This compound has a quasi-one-dimensional quarter-filled
$\pi$-band system. $\rm{(DI-DCNQI)_{2}Ag}$ undergoes a transition to a charge-ordered
insulator, where the 3:1 charge disproportionation occurs on alternate molecules
within a chain. This compound shows continuous and first-order phase boundaries
together with TCP on the phase boundary between metals and charge order
drawn in the phase diagram in the plane of temperature and pressure.
The pressure is supposed to control the bandwidth $t$ in our model
and the extended Hubbard model (\ref{EHM}) offers a minimal model and
may be a good starting point to discuss the qualitative aspects of
the phase diagram as well as the critical behavior.
The three dimensional charge order takes place in this compound.
The upper critical dimension of TCP is three, so it is expected that
the singularity of conductivity is well described by that of the Hartree-Fock approximation.  
The universality and critical exponents at charge-order transitions
should not depend on details of the charge-order structure and the estimated
exponents in the extended Hubbard model in the previous sections must apply here.

\begin{figure}[htp]
	\begin{center}
		\includegraphics[width=6cm]{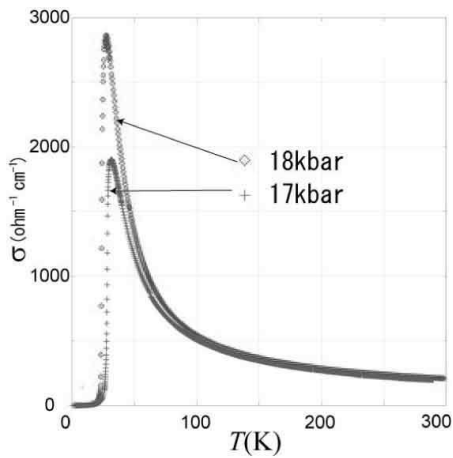}   
	\end{center}
\caption{Experimental data of the conductivity near TCP~\cite{DCNQI}.}
\label{fig:Ex_Con}
\end{figure}

Resistivity has been measured near TCP by Itou $et$ $al.$~\cite{DCNQI} and they
have estimated that the location of TCP is near 18 kbar. We analyze the singularity of conductivity
at 18 kbar and 17 kbar. Experimental data of the conductivity is shown in Fig. \ref{fig:Ex_Con}.      

\begin{figure}[htp]
	\begin{center}
		\includegraphics[width=5cm]{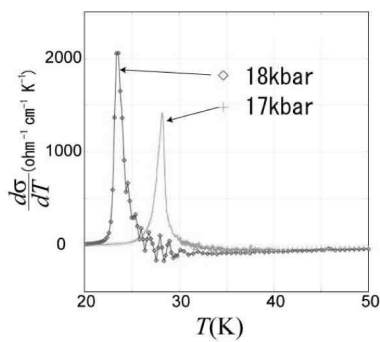}   
	\end{center}
\caption{Experimental data of $d\sigma/dT$ near TCP~\cite{DCNQI}.}
\label{fig:Ex_Dif_Con}
\end{figure}

 We differentiate the conductivity with respect to temperature $T$.
If the conductivity has a singularity such as 
$\sigma-\sigma_{c}\sim|T-T_{c}|^{q}(A\log{|T-T_{c}|}+B)$,
the differentiation of the conductivity behaves as 
$d\sigma/dT \propto |T-T_{c}|^{q-1}\log{|T-T_{c}|}$.
If $q$ is less than 1, $d\sigma/dT$ should diverge at $T_{c}$.
Even if $q$ is equal to 1, $d\sigma/dT$  has logarithmic divergence.
Data of $d\sigma/dT$ is shown in Fig. \ref{fig:Ex_Dif_Con}.

 We assign the critical temperature $T_{c}$ as the middle point
between the highest and the second highest points of $d\sigma/dT$.
Using this $T_{c}$, we determine $\sigma_{c}$, and perform
scaling analysis. We estimate $T_{c}=28.23$ K, $\sigma_{c}=1260.13$ohm$^{-1}$cm$^{-1}$
for 17 kbar and $T_{c}=23.49$ K, $\sigma_{c}=992.37$ohm$^{-1}$cm$^{-1}$ for 18 kbar.

\begin{figure}[htp]
	\begin{center}
		\includegraphics[width=6cm]{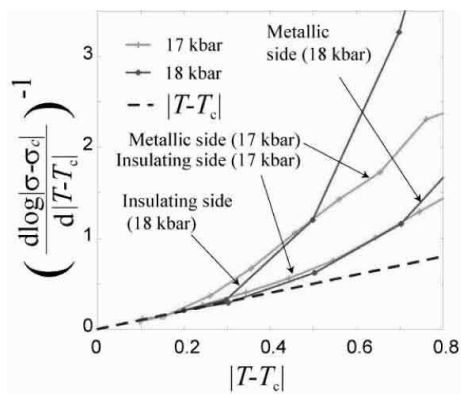}   
	\end{center}
\caption{Scaling analysis of $\sigma$ at 17 kbar and 18kbar in the metallic 
and insulating sides. The slope of the broken straight line 
without symbols corresponds 
to the critical exponent of conductivity $q=p_{t}^{\prime}=1$
without logarithmic correction.
We estimate slope as $1.03(6)$ for 17 kbar and $1.1(1)$. }
\label{fig:Dif_Con_Ex}
\end{figure}

From the analysis of Hartree-Fock approximation,
if the singularity of conductivity comes from 
that of the carrier density, 
it is expected  that conductivity has the singularity 
$|\sigma-\sigma_{c}|\propto |T-T_{c}|^{q}(A\log|T-T_{c}|+B)$,
where $q=p_{t}=1/2$($q=p_{t}^{\prime}=1$) 
in the canonical (grand-canonical) ensemble, respectively. 
To estimate $q$, we plot 
\begin{align}
	S=\left(\frac{d\log|\sigma-\sigma_{c}|}{d|T-T_{c}|}\right)^{-1}
	=|T-T_{c}| \notag \\
	\times\left(q+\frac{1}{A\log|T-T_{c}|+B}\right)^{-1}.
\end{align}
We note that the slope of $S$ becomes $q^{-1}$ in the region where $|T-T_{c}|\ll1$.  
The results of scaling analysis are shown in Fig. \ref{fig:Dif_Con_Ex}.

In both data for 17 and 18 kbar, slopes become $1$ near $T_{c}$. 
These results indicate that the phase separation occurs along the first-order transition line
and terminates at TCP in $\rm{(DI-DCNQI)_{2}Ag}$. In this case, the Hartree-Fock analysis 
in the grand-canonical ensemble is justified in three dimensions
and the critical exponent is $q=p_{t}^{\prime}=1$.
Therefore the experimental data for 17 and 18 kbar are sufficiently close 
to the critical region of TCP in the grand-canonical ensemble. 
We do not exclude another possibility 
that the phase separation does not occur
or terminates on the first-order transition line 
of the canonical ensemble in $\rm{(DI-DCNQI)_{2}Ag}$.
In this case, Hartree-Fock analysis in the canonical 
ensemble is justified in three dimensions and the critical exponent is $q=p_{t}=1/2$. 
Therefore, the experimental data are not sufficiently close to the critical 
region of TCP. If one approaches the critical region of TCP,
the singularity of the conductivity should approach $q=p_{t}=1/2$
if the phase separation is absent.
More complete understanding is left for future studies.
It is desired to measure the critical properties of the conductivity
in the region closer to TCP after identifying TCP precisely.
It is also intriguing to observe the degrees of carrier-density
inhomogeneity in samples, which is in fact helpful 
in revealing the tendency for the phase separation.

Here, we consider the singularity of relaxation time.
Phenomelogically, carrier density $X$ (relaxation time
$\tau$) is divided into the coherent part $X_{coh}$($\tau_{coh}$) and 
the incoherent part $X_{inc}$($\tau_{inc}$).
Then, the conductivity may be expressed as 
$\sigma\sim X_{coh}\tau_{coh}+X_{inc}\tau_{inc}$.
If we consider the singular part of $X_{coh}$($X_{inc}$) and $\tau_{coh}$($\tau_{inc}$)
associated with the transition as $X_{coh}^{s}$($X_{inc}^{s}$) and $\tau_{coh}^{s}$($\tau_{inc}^{s}$),
such that $X_{coh}=X_{coh}^{0}+X_{coh}^{s}$($X_{inc}=X_{inc}^{0}+X_{inc}^{s}$),
$\tau_{coh}=\tau_{coh}^{0}+\tau_{coh}^{s}$($\tau_{inc}^{s}=\tau_{inc}^{0}+\tau_{inc}^{s}$)
with the normal parts $X_{coh}^{0}$($X_{inc}^{0}$) and $\tau_{coh}^{0}$($\tau_{inc}^{0}$),
we obtain $X_{coh}\tau_{coh}\sim X_{coh}^{0}\tau_{coh}^{s}+X_{coh}^{s}\tau_{coh}^{0}+X_{coh}^{s}\tau_{coh}^{s}$
($X_{inc}\tau_{inc}\sim X_{inc}^{0}\tau_{inc}^{s}+X_{inc}^{s}\tau_{inc}^{0}+X_{inc}^{s}\tau_{inc}^{s}$), respectively.
First, we consider the singularity of the coherent part.
It is expected that $X_{coh}^{s}$ has the same singularity as
that of $X$. Namely, the singularity of $X_{coh}^{s}$
is given as $|T-T_{c}|^{p_{t}}(A\log{|T-T_{c}|}+B)$.
In Appendix C, it is speculated that $\tau_{coh}$ either does not have 
the normal part $\tau_{coh}^{0}$ or does have the nonzero $\tau_{coh}^{0}$.
Here, $\tau_{coh}^{s}$ has the same singularity as that of the carrier density $X$
except for logarithmic correction.
From this, the dominant singularity of $X_{coh}\tau_{coh}$
comes from  $\tau_{coh}^{s}$ or $X_{coh}^{s}$ and the singularity of $\tau_{coh}^{s}$
is weaker than that of the carrier density.
Next, we consider the singularity of the incoherent part.
Again, it is expected that $X_{inc}^{s}$ has the same singularity as
that of $X$. One of the evidences that $\tau_{inc}^{s}$ has a weaker singularity
than that of the carrier density is found in the study of the double-exchange model
which describes ferromagnetic transition in perovskite-type manganese oxides.
Theoretically, the singularity of the relaxation time near the ferromagnetic transition
is studied in the double-exchange model within the mean-field framework~\cite{Kubo,Furukawa}. 
In this system, carrier density does not change at the ferromagnetic transition,
and the singularity of the conductivity comes from that of the relaxation time.
The singularity of the relaxation time $\tau_{inc}^{s}$ is characterized as 
that of the square of the order parameter $m$ (spontaneous magnetization).
Therefore, $\tau_{inc}^{s}$ is given as
\begin{equation}
	\tau_{inc}^{s}\propto m^{2}\propto |T-T_{c}|^{2\beta}.
\end{equation}
Experimentally, for example in La$_{1-x}$Sr$_{x}$MnO$_{3}$, 
the singularity of the resistivity is also  characterized as
that of  $m^{2}$ near the transition point~\cite{Tokura}.
This singularity is weaker than that of the carrier density
given in eq. (\ref{eq:Log_correct}).
This argument is for the case of ferromagnetic transition
and it is not clear how it is universal.
However it is likely that in the case of the charge-order system
it shares the similar mechanism and
$\tau_{inc}^{s}$ has the similar singularity.

We finally obtain the singularity of $\sigma$
as the same one with $X$.
Detail analysis of the singularity of the
relaxation time in the charge-order system
is left for future studies.
        
\section{Summary and Discussion}
\label{Summary}
A simple charge-order transition has been studied as a typical example to understand the 
common and generic feature.
An important aspect of the transition is that it has both continuous and first-order
transitions connected by the tricritical point (TCP).
We have studied charge-order transitions in the extended Hubbard model.
By introducing the onsite interaction $U$ and the nearest-neighbor repulsion $V$,
this model shows the transition to a charge-ordered phase at low temperatures
with alternating charge density for 
the bipartite lattice, when $V$ increases.
In the classical limit of ignoring the itinerancy, the model has been solved 
first by the mean-field approximation and then on a square lattice 
by Monte-Carlo simulation with the exchange Monte Carlo algorithm 
to get around the critical slowing down.

Our results for critical exponents agree with those in the literature for the 
available studied quantities.
On the lambda line, it reproduces the Ising exponents, $\eta=0.25$, $\nu=1.0$ and 
logarithmic divergence of the specific heat,
which are consistent with Onsager's exact solution of the 2D Ising model.
Using Ehrenfest law, we have shown that this divergence of specific heat causes 
the logarithmic divergence of the doublon susceptibility $\chi_{D(\bf{0})}$,
the susceptibility of the nearest-neighbor charge correlation $\chi_{R}$ and
the charge susceptibility $\chi_{c}$ on the lambda line. 
At the TCP, we obtain the order parameter exponents, $\eta_t=0.11\pm0.02$, $\nu_t=0.55\pm 0.01$ while
for the susceptibility of the order parameter exponent and
the doublon susceptibility exponent defined in eq. (\ref{eq:def_gamma_t}) and eq. (\ref{eq:def_gamma_2t}),
we obtain $\gamma_{t}=1.04\pm0.06$ and
$\gamma_{2t}=0.90\pm0.06$, respectively. 
Furthermore we clarify which fluctuations diverge at TCP both 
in the mean-field approximation and Monte Carlo calculations.
For the mean-field approximation and the Monte Carlo calculations,
we have shown that, the specific heat $C$, the susceptibility of the nearest-neighbor charge correlation $\chi_{R}$ and
the charge susceptibility $\chi_{c}$ diverge with the same exponent as that of
 $\chi_{D(\bf{0})}$, namely $\gamma_{2t}$ at TCP.
However $\chi_{c}$ has a subtlety.
Because of the particle-hole symmetry, 
$\chi_{c}$ does not diverge at half filling,
while, away from half filling, we confirm that $\chi_{c}$ diverges at TCP. 
As far as the authors know, the divergence of $\chi_{c}$ at TCP has not been recognized in the literature. 
The comparison of the mean-field results with the Monte Carlo results 
indicate that the both results show the divergences in the same quantities.   
Exact results by Monte Carlo calculations give
 only quantitative modifications of the critical exponents
from the mean-field results.

In the itinerant model, within the Hartree-Fock study, we obtain TCP at a finite temperature
for $t^{\prime}/t=0.2$. At TCP, we obtain the critical exponent of the order parameter $\beta_{t}=1/4$. 
We show that the doublon susceptibility $\chi_{D(\bf{0})}$, and 
the susceptibility of the nearest-neighbor charge correlation $\chi_{R}$ diverge with 
the singularity $\sim|g-g_{c}|^{-\gamma_{2t}}$ at TCP.
Here, $\gamma_{2t}$ is equal to the classical mean-field value $1/2$. 
We also show that the charge susceptibility $\chi_{c}$ diverges at TCP in the
grand-canonical ensemble  with  the singularity $\sim|g-g_{c}|^{-\gamma_{2}}$
where $\gamma_{2}=1$.
The overall exponents are the same as those in the classical model.

 At the TCP, various physical quantities
diverge much more strongly than those at the lambda transition. 
In particular, diverging charge fluctuations at zero wavenumber
revealed around TCP may mediate various instabilities toward
ordering such as superconductivity,
when the divergences are involved in the Fermi degeneracy
with Fermi surface instability.
The consequence of such strong divergences clarified at the
TCPs will be discussed in a separate publication.

We have also obtained 
the conductivity exponent, which is specific to the itinerant model. 
In three-dimensional systems, these exponents including the conductivity exponent may be 
correct because the upper critical dimension is three, while in two dimensions, 
the exponents may be modified as in the classical model, which we have shown in the Monte Carlo result.
The conductivity exponent beyond the mean-field level in two-dimensional systems is left for 
future studies. 

On the conductivity, we obtain the exponent $p_{t}$ 
which is defined by $|X-X_{c}|=|g-g_{c}|^{p_{t}}(A\log|g-g_{c}|+B)$ in the canonical ensemble at TCP. 
The exponent $p_{t}$ has a relation to the order parameter
exponent $\beta_{t}$ as $p_{t}=2\beta_{t}=1/2$. 
We also show that $p_{t}$ changes into $p_{t}^{\prime}=2p_{t}=1$ at TCP in the
grand-canonical ensemble when TCP in the canonical ensemble is in the phase separation region.
Similarly, on the lambda line, the exponent defined by $|X-X_{c}|=|g-g_{c}|^{p}(A\log|g-g_{c}|+B)$ 
is given by $p=2\beta=1$ on the lambda line, where $p$ should be $2p_{t}$. 

The experimental data for (DI-DCNQI)$_2$Ag~\cite{DCNQI} show $q\sim1$ for the points
closest to the tricriticality, where $q$ is defined from the conductivity 
$\sigma\propto (g-g_c)^{q}(A\log|g-g_{c}|+B)$.
There are two possibilities for explaining the experimental results. One is that the phase separation
occurs and  terminates at TCP in (DI-DCNQI)$_2$Ag. 
In this case, the conductivity should have the same singularity as 
that of the Hartree-Fock analysis
in the grand-canonical ensemble, namely, $p_{t}^{\prime}=1$, which is consistent with
the experimental value $q\sim1$. 
Even when the long-ranged Coulomb repulsion prohibits the real
phase separation, it is conceivable that the formation
of microdomains results in effectively similar exponents in experiments.
Another possibility is that the phase separation does not
occur and terminates on the first-order transition line in (DI-DCNQI)$_2$Ag. 
In this case, the conductivity should have the same singularity as that of
the  Hartree-Fock analysis in the canonical ensemble, namely $p_{t}=1/2$.
The experimental results are not consistent with our Hartree-Fock value.
Then the origin of this discrepancy must be ascribed to the interpretation
that the experimental data is not sufficiently close to the critical region of TCP.
If one approaches the critical region of TCP, $q$ should be close to $p_{t}=1/2$.
To clarify the origin of singularity,  
it would be desired to perform more detailed experimental studies.   

We have discussed quantum effects through the singularity of the conductivity,
while the presence of the transfer does not alter the exponents for the order parameter $\Delta$ 
itself from the classical value. This may be due to the fact that the transition temperature
is sufficiently high so that the quantum proximity is not visible for the charge-order transition itself.
If the tricritical temperature could be lowered, its quantum effect would be observed. 
However, in the extended Hubbard model, the tricritical line terminates at a finite temperature
and cannot be lowered to zero as one sees in the schematic phase diagram in Fig. \ref{fig:G_Phase}.
The quantum effect on TCP would be an intriguing issue to be studied in a different situation in the future.

In the itinerant model, within the Hartree-Fock study, we have ignored the possibility of the 
magnetic order.
However, antiferromagnetic and metal-insulator transitions of the extended Hubbard model with $V=0$
can also be studied in the framework by putting $V=0$, where the interaction effect from $U$
is only reflected in $g$ in our treatments. Then the same argument for the criticalities apply in the region
of $g<0$, because transforming the charge order to the antiferromagnetic order with the transformation
of $g$ to $-g$ leaves the self-consistent equation unchanged.
This means that the Hubbard model with only the onsite repulsion $U$ with the transfers $t$ and $t'$
has the same phase diagram and criticalities simply by replacing the charge order with the antiferromagnetic
order. This is a one-to-one equivalence within the Hartree-Fock approximation.
However, in the true phase diagram of the Hubbard model, it has a 
subtlety because the metal-insulator
boundary (Mott transition) may extend beyond the antiferromagnetic boundary contrary to the artifact of the
Hartree-Fock results, while in case of the charge 
order, the insulating phase without the charge order should not exist and the Hartree-Fock phase diagram
is correct in this aspect.

We find that the tricritical line terminates when $t'/t$ is increased and the phase boundary of 
metal-insulator transition separates from that of the charge-order transition.
This generates an intriguing first-order metal-insulator boundary with marginally quantum critical 
behavior~\cite{MQMCP} within the mean-field level. This issue will be discussed in a separate paper.

\section*{Acknowledgements}
The authors would like to thank K. Kanoda and T. Itou for sending us 
their experimental data and useful discussions.
One of the authors(T. M.)  would like to thank S. Watanabe for stimulating discussions. 
This work has been supported from the Grant in Aid for Scientific Research
on Priority Area  from the Ministry of Education,
Culture Sports, Science and Technology.  
A part of our computation in this work has been done using the
facilities of the Supercomputer Center, Institute for Solid
State Physics, University of Tokyo.

\section*{Appendix A: Ehrenfest law}
\begin{figure}[htp]
	\begin{center}
		\includegraphics[width=6cm]{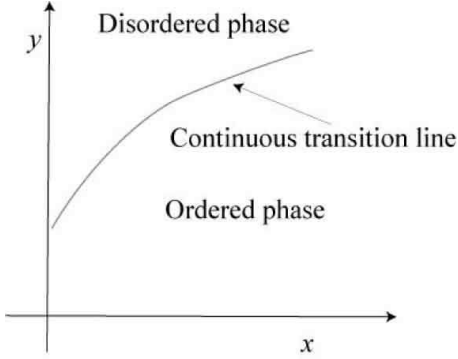}   
	\end{center}
\caption{Continuous transition line in  two-dimensional plane.}
\label{fig:Ehren}
\end{figure}

	We explain the general framework of Ehrenfest law in this appendix~\cite{Thermo}.
We consider the situation that  the continuous transition line lies in a two-dimensional
plane in a set of parameter space (see Fig. \ref{fig:Ehren}).
We consider two physical properties $X$ and $Y$, which are conjugate to $x$ and $y$, respectively. 
We define $X$ and $Y$ as the first 
derivatives of the free energy $F$ with respect to $x$ and $y$ as

\begin{align} 
	X&=\frac{\partial F}{\partial x},Y=\frac{\partial F}{\partial y}, \notag \\
\chi_{X}&=\frac{\partial X}{\partial x},\chi_{Y}=\frac{\partial Y}{\partial y}. 
\end{align} 

We assume that $X$ and $Y$ are continuous along the continuous transition line,
but the derivatives of $X$ and $Y$ are not continuous. Then we define two limits
of the second derivatives as
\begin{align}
\frac{\partial X}{\partial x}|_{dis}&=
\lim_{x\to x_{c}^{+},y\to y_{c}^{+}}\frac{\partial X}{\partial x}, \notag \\
\frac{\partial X}{\partial x}|_{order}&=
\lim_{x\to x_{c}^{-},y\to y_{c}^{-}}\frac{\partial X}{\partial x}, 
\end{align}
where the subscripts $dis$ and $order$ express the derivatives
in the disordered and ordered sides at the transition line,
respectively.
 
$X$ is expanded by $x$ and $y$ near a critical point $(x_c,y_c)$ as 
\begin{align}
&X(x_c+\delta x,y_c+\delta y)=X(x_c,y_c) \notag \\
&+\frac{\partial X}{\partial x}|_{dis}\delta x 
+\frac{\partial X}{\partial y}|_{dis}\delta y
+(\delta x^2,\delta y^2)    :\text{disorder side}, \notag \\
&X(x_c+\delta x,y_c+\delta y)=X(x_c,y_c) \notag \\
&+\frac{\partial X}{\partial x}|_{order}\delta x
+\frac{\partial X}{\partial y}|_{order}\delta y
+(\delta x^2,\delta y^2)    :\text{order side}. 
\end{align}
Along the continuous line, $X$ at the disorder and order phases have
the same value at $(x_c,y_c) $. We obtain the relation 
\begin{align}
	X(x+\delta x,y+\delta y)_{dis}-X(x+\delta x,y+\delta y)_{order}=0.
\end{align}
From this relation, we obtain the slope of the continuous transition 
line.
\begin{align}
	\frac{\delta y}{\delta x}&=-\frac{\frac{\partial X}{\partial x}|_{dis}-\frac{\partial X}{\partial x}|_{order} }
{\frac{\partial X}{\partial y}|_{dis}-\frac{\partial X}{\partial y}|_{order}} \notag \\
	&=-\frac{\chi_{X}|_{dis}-\chi_{X}|_{order} }
{\frac{\partial X}{\partial y}|_{dis}-\frac{\partial X}{\partial y}|_{order}} \label{Ehren_1} 
\end{align}
Using the same discussion for $Y$, we obtain the relation as 
\begin{align}
	\frac{\delta y}{\delta x}&=-\frac{\frac{\partial Y}{\partial x}|_{dis}-\frac{\partial Y}{\partial x}|_{order} }
{\frac{\partial Y}{\partial y}|_{dis}-\frac{\partial Y}{\partial y}|_{order}} \notag \\
	&=-\frac{\frac{\partial Y}{\partial x}|_{dis}-\frac{\partial Y}{\partial x}|_{order}}
{\chi_{Y}|_{dis}-\chi_{Y}|_{order} }. \label{Ehren_2} 
\end{align}

Along the continuous $``$line$"$, second partial differentiations are well defined.
Thus, the relations  such that $\frac{\partial X}{\partial y}|_{dis}=\frac{\partial Y}{\partial x}|_{dis}$ ,
$\frac{\partial X}{\partial y}|_{order}=\frac{\partial Y}{\partial x}|_{order}$ are satisfied.

We assume that $\frac{\delta x}{\delta y}$ is finite.
In this condition, if $\chi_{X}$ diverges, $\frac{\partial X}{\partial  y}$ should diverge. Then,
from the relation $\frac{\partial X}{\partial y}=\frac{\partial Y}{\partial x}$, $\frac{\partial X}{\partial y}$ should diverge.
Finally, $\chi_{Y}$ also diverges with the same singularity of $\chi_{X}$
and vice versa. This is the Ehrenfest law.

If we take $X$ as 
the onsite interaction $U$ and $Y$ as
the temperature $T$,
$\chi_{X}$ and $\chi_{Y}$ correspond to 
the doublon susceptibility $\chi_{D({\bf 0})}$
and the specific heat $C$, respectively.
From this, if the slope of the continuous transition
line is finite,  $\chi_{D({\bf 0})}$ has the
same singularity as that of $C$.

We summarize the relations with $\chi_{X}$ or $\chi_{Y}$ and physical properties below. 
\begin{align}
	\chi_{Y}&=C \ \ \text{($Y=T$)} \label{eq:AP_C}\\
	\chi_{X}&=\chi_{D({\bf 0})} \ \  \text{($X=U$)} \label{eq:AP_D} \\
	\chi_{X}&=\chi_{R} \ \  \text{($X=V$)} \label{eq:AP_R} \\
	\chi_{X}&=\chi_{c} \ \  \text{($X=\mu$)} \label{eq:AP_c}
\end{align}

\section*{Appendix B: Singularity of carrier density}

In this appendix, we obtain the singularity of the carrier
density $X$ defined in eq. (\ref{eq:Carrier_Density}). We evaluate $X$ near $(\pi/2, \pi/2)$
and obtain logarithmic correction.

Near $(\pi/2, \pi/2)$, upper band $E_{+}$ is approximated as
\begin{align}
	E_{+}&\sim4t^{\prime}\delta x\delta y+\sqrt{4t^{2}(\delta x+\delta y)^{2}+\Delta^{2}} \\
		 &=t^{\prime}(\alpha_{1}^{2}-\alpha_{2}^{2})+\sqrt{4t^{2}(\alpha_{1})^{2}+\Delta^{2}},	
\end{align}
where we define $\delta x$[$\delta y$] and $\alpha_{1}$[$\alpha_{2}$]
as $\delta x=\pi/2-k_{x}$[$\delta y=\pi/2-k_{y}$] and $\alpha_{1}=\delta x+\delta y$
[$\alpha_{2}=\delta x-\delta y$], respectively.

Using this band dispersion, the singular part of $X$ is approximated as
\begin{equation}
\tilde{X}=\int \frac{d\alpha_{1}d\alpha_{2}}
{1+\exp{\beta[t^{\prime}(\alpha_{1}^{2}-\alpha_{2}^{2})+\sqrt{4t^{2}(\alpha_{1})^{2}+\Delta^{2}}-\mu]}}.
\end{equation}
Near the transition point,  
chemical potential can be expanded with respect to $\Delta$ as
\begin{equation}
	\mu=\mu_{0}+\mu_{1}\Delta^{2},
\end{equation}
where $\mu_{0}$ and $\mu_{1}$ are constants. 

To obtain a stronger singularity than that of $\Delta^{2}$,
we consider $\partial \tilde{X}/\partial \Delta^{2}$.  
\begin{align}
 \frac{\partial \tilde{X}}{\partial \Delta^{2}}&=\int d\alpha_{1}d\alpha_{2}g(\beta,E_{+}) \notag \\
 &\times{[\frac{1}{\sqrt{4t^{2}\alpha_{1}^{2}+\Delta^{2}}}-\mu_{1}]}, \label{eq:Log}
\end{align}
where $g(\beta,E_{+})=-\beta e^{\beta(E_{+}-\mu)}/(1+e^{\beta(E_{+}-\mu)})^{2}$.
Since $g(\beta,E_{+})$ does not diverge at a finite temperature,
its integration should be $O(1)$. 
Therefore, near the transition point ($\Delta\ll 1$),
 the strongest singularity of eq. (\ref{eq:Log}) comes from  
$\int d{\alpha_{1}}/{\sqrt{4t^{2}\alpha_{1}^{2}+\Delta^{2}}}$.
Then we obtain the singularity of $\tilde{X}$ as 
\begin{align}
	\tilde{X}&\propto\int_{0}^{\Lambda} d\alpha_{1}\frac{1}{\sqrt{4t^{2}\alpha_{1}^{2}+\Delta^{2}}} \label{eq:Log_2} \\ 
	 &=\Bigl[\log[{\alpha_{1}+\sqrt{\alpha_{2}^{2}+(\Delta/2t)^{2}}}]/(2t)\Bigr]^{\Lambda}_{0}	\\
     &\sim\log[{|\Delta|/2t}]/t,
\end{align} 
where $\Lambda$ is a cutoff parameter.
From this, we obtain the singularity of the carrier density $X$ as
\begin{eqnarray}
	X\propto\Delta^{2}(A\log{\Delta}+B),\label{eq:Log_correct}
\end{eqnarray}
where $A$ and $B$ are constants.

Here, we note that, since the logarithmic correction comes from eq. (\ref{eq:Log_2}),
this logarithmic correction appears in any dimensions.  

\section*{Appendix C: Singularity of relaxation time}
In this appendix, we analyze the singularity of
the coherent part of the relaxation time $\tau_{coh}$ near TCP
following the conventional Fermi-liquid scheme~\cite{RMP,Fermi}.
In this scheme, although the singularity of $\tau_{coh}$ is caused by
the antiferromagnetic spin fluctuation, it is likely
that the order-parameter fluctuation of the charge order
causes a similar singularity. 

We now consider the singularity taking place
from the disordered (non-charge-ordered) phase.
The relaxation time $\tau_{coh}$ is proportional to
the inverse of the imaginary part of self-energy ${\rm Im}\Sigma$.
In the Fermi-liquid scheme, ${\rm Im}\Sigma$ is approximated as
\begin{equation}
	{\rm Im}\Sigma\sim [\omega^{2}+(\pi T)^{2}]\sum_{\bf k}
	\Gamma_{\uparrow\downarrow}({\bf k})^{2},
\end{equation}
where $\Gamma_{\uparrow\downarrow}({\bf k})$ is the
four-point vertex function. The four-point vertex function 
$\Gamma_{\uparrow\downarrow}({\bf k})$ has the relation
with the susceptibility $\chi({\bf k})$ such that
\begin{equation}
	\Gamma_{\uparrow\downarrow}({\bf k})
\sim \frac{\chi({\bf k})}{\chi_{b}({\bf k})^{2}}, 
\end{equation}
where $\chi_{b}({\bf k})$ is a noninteracting
susceptibility. Near the transition point, $\chi({\bf Q+q})$
can be approximated as
\begin{equation}
	\chi({\bf Q+q})\sim\frac{1}{K^{2}+q^{2}}.
\end{equation}
Using this relation, we obtain the singularity of Im$\Sigma$ as
\begin{align}
	{\rm Im}\Sigma\propto\sum_{\bf k}
	\Gamma_{\uparrow\downarrow}({\bf k})^{2} 
	\propto \int dk \frac{k^{d-1}}{(K^{2}+k^{2})^{2}}\propto K^{d-4}.
\end{align}
In a three-dimensional system, near TCP in the canonical ensemble,
 $\chi(\bf{Q})$ diverges as
\begin{equation}
	\chi({\bf Q})\propto |T-T_{c}|^{-\gamma_{t}}.
\end{equation}
From this, the singularity of $K$ is given as
\begin{equation}
	K\propto |T-T_{c}|^{\gamma_{t}/2},
\end{equation} 
where $\gamma_{t}=1$.
Therefore in a three-dimensional system, the singularity
of $\tau_{coh}$ is obtained as
\begin{equation}
	\tau_{coh}\propto \tau_{coh}^{s}=({\rm Im}\Sigma)^{-1}\propto|T-T_{c}|^{1/2}.
\end{equation}
Here, if we employ the above analysis, $\tau_{coh}$ has no normal part $\tau_{coh}^{0}$.
On the other hand, near TCP in the grand-canonical ensemble, $\gamma_{t}=1$ changes
into $\gamma=2$. Therefore, the singularity
of $\tau_{coh}$ changes into
\begin{equation}
	\tau_{coh}\propto \tau_{coh}^{s}=|T-T_{c}|.
\end{equation}
It should be noted that $\tau_{coh}$ has the same
singularity as that of the carrier density $X$
except logarithmic correction
in both the canonical and grand-canonical ensembles.
The singularity of $X$ is logarithmically stronger
than that of $\tau_{coh}$. 

We, however, also note that the above analysis would
oversimplify the real situation, where the momentum dependence
of the relaxation time ignored in the above analysis could
become important. In fact, if the pocket-type Fermi surface
is expected in the ordered phase,
strongly anisotropic arc-type structure
may appear in the disordered phase. 
Then coherent quasiparticles on the arc along the Fermi surface 
cannot be scattered to other arc point 
by the scattering wave vector $(\pi,\pi)$.
The diverging fluctuation at $(\pi,\pi)$ controls 
the quasiparticle relaxation time, and the $(\pi,\pi)$ vector connects an arc point
to the missing counterpart of the pocket,
while such a missing part does not have a well-defined Fermi surface
and such scattering process is not effective.
If this circumstance applies, $\tau$ remains nonsingular
and nonzero. Then we expect $\tau$ to be expressed by
$\tau=\tau_{0}+\tau_{s}$, where $\tau_{0}$ is the
remaining regular part.
We also expect a similar anisotropy in the
self-energy in the ordered phase.


\begin{thebibliography}{99}
\bibitem{RMP}For  review, see M. Imada, A. Fujimori and Y. Tokura: Rev. Mod. Phys. {\bf 70} (1998) 1039.
\bibitem{Seo}
H. Seo, C. Hotta, and H. Fukuyama: Chem. Rev. {\bf104} (2004) 5005.
\bibitem{Noda}Y. Noda and M. Imada:
Phys. Rev. Lett. {\bf 89} (2002) 176803
\bibitem{DCNQI}
T. Itou, K. Kanoda, K. Murata, T. Matsumoto
, K. Hirai, and T. Takahashi: Phys. Rev. Lett. {\bf 93}  (2004) 216408.
\bibitem{TCP}
For a review, see I. D. Lawrie and S. Sarbach: $Phase$
$Transition$ $and$ $Critical$ $Phenomena$, Vol. 9, p. 2 eds.
 C. Domb and J. L. Lebowitz  (Academic Press, London, 1984)
\bibitem{BEG}
M. Blume, V. J. Emery, R. B. Griffiths: Phys. Rev. A {\bf 4} (1971) 1071
\bibitem{Nigel}
For example, see N. G. Nigel: $Lectures$ $on$ $Phase$ $Transitions$ $and$ $Renormalization$ $Group$
(Addison-Wisely, 1992) 
\bibitem{MC}
D. P. Landau and K. Binder: $A$ $Guide$ $to$ $Monte$ $Carlo$ $Simulations$
$in$ $Statistical$ $Physics$ (Cambridge University Press, Cambridge, 2000) 
\bibitem{EXMC}
K. Hukushima and K. Nemoto: J. Phys. Soc. Jpn. {\bf 65} (1995) 1604
\bibitem{MCRG}
D. P. Landau and R. H. Swendsen: Phys. Rev. Lett. {\bf 46} (1981) 1437
\bibitem{Kaufmann}
B. Kaufmann and L. Onsager: Phys. Rev. {\bf 76} (1944) 1244
\bibitem{Thermo}
S. Watanabe and M. Imada: J. Phys. Soc. Jpn. {\bf 73} (2004) 3341
\bibitem{DCNQI-NMR}
K. Hiraki and K. Kanoda: Phys. Rev. Lett. {\bf 80} (1998) 4737
\bibitem{MQMCP}
M. Imada: Phys. Rev. B {\bf 72} (2005) 075113; J. Phys. Soc. Jpn. {\bf 74} (2005) 859
\bibitem{Kubo}
K. Kubo and N. Ohata: J. Phys. Soc. Jpn. {\bf 33} (1972) 21
\bibitem{Furukawa}
N. Furukawa:  J. Phys. Soc. Jpn. {\bf 63} (1994) 3214
\bibitem{Tokura}
Y. Tokura, A. Urushibara, Y. Moritomo,
T. Arima, A. Asamitsu, G. Kido and N. Furukawa:
 J. Phys. Soc. Jpn. {\bf 63} (1994) 3931
\bibitem{Fermi}
H. Kohno and K. Yamada: Prog. Theor. Phys. {\bf 85} (1991) 13

\end{thebibliography}
\end{document}